\documentclass[proof]{WileyASNA-v1}
\usepackage{gensymb}

\usepackage[section]{placeins}
\usepackage[protrusion=true, expansion=true, shrink=55, stretch=55, 
tracking=true, kerning=true, spacing=true, 
final]{microtype}
\usepackage{caption}
\DeclareCaptionLabelFormat{abbreviated}{#1~#2 (abbreviated)}

\articletype{Article Type}%

\received{3 March 2026}
\revised{21 July 2026}
\accepted{28 July 2026}

\newcommand*\degr{\ensuremath{^\circ}}
\newcommand*\arcmin{\ensuremath{^\prime}}
\newcommand*\arcsec{\ensuremath{^{\prime\prime}}}
\def\utw{\ensuremath{\smash{\rlap{\lower5pt\hbox{$\sim$}}}}}
\def\udtw{\ensuremath{\smash{\rlap{\lower6pt\hbox{$\approx$}}}}}

\newcommand*\farcm{\ensuremath{\overset{\prime}{.}}}
\newcommand*\farcs{\ensuremath{\overset{\prime\prime}{.}}}

\newcommand\micron{\mbox{$\mu$m}}%

\begin{document}

\title{A Citizen Science Search for Compact Emission Line Nebulae}

\author[1]{Bringfried Stecklum*}

\author[2]{Peter Bresseler}

\authormark{B. Stecklum \& P. Bresseler}

\address[1]{\orgname{Thüringer Landessternwarte, Sternwarte 5, 07778 Tautenburg}, \country{Germany}}

\address[2]{\orgname{Auf der Strenge 38, 22397 Hamburg \country{Germany}}}

\corres{
\email{stecklum@tls-tautenburg.de}}

\presentaddress{Sternwarte 5, 07778 Tautenburg, Germany}

\abstract{In recent years, citizen science became an integral part of astronomy. Numerous projects contribute to important studies by involving large communities. While these are getting quite some attention, there are also individual efforts, which are nevertheless very valuable. Here, we report on results of a search for compact optical emission line nebulae, which are signposts of both young and evolved stars. The increased brightness in  red photometric bands due to the contribution of shock-excited emission lines has been used
as a tracer when examining optical broadband surveys. With the help of public databases, we identified and characterized associated sources if possible. Most of them are young stellar objects (YSOs). }

\keywords{surveys, ISM: jets and outflows, Herbig-Haro objects, stars: formation, planetary nebulae: general }


\maketitle


\section{Introduction}\label{sec1}

Compact optical emission line nebulae are a common feature of YSOs as well as proto-planetary nebulae (PPN), despite their completely different evolutionary stages, marking the beginning and the near end of the stellar life cycle. For YSOs, these nebulae, dubbed Herbig-Haro objects (HHOs), result from collisional excitation due to shocks of jets with internal or ambient matter (\citealt{1951ApJ...113..697H}; \citealt{1952ApJ...115..572H}). The most famous example is HH30 \citep{1974LicOB.658....1H}, which appeared as a compact nebulosity in ground-based imaging but was resolved by the Hubble Space Telescope to be an edge-on protoplanetary disk surrounding a protostar that drives a bipolar jet \citep{1996ApJ...473..437B}. Thus, HHOs are clear signposts of star formation. 

Our successful search for compact emission nebulae led to the discovery of Bres 1 \citep{2023RNAAS...7..254B}, a YSO associated with a probable Herbig-Haro object.  In the meantime, we have identified more similar faint nebulosities by careful and laborious examination of available online images from various sky surveys, which are reported here.

\section{Data acquisition and analysis}\label{sec2}
The primary tool for the search and identification of these objects is the Aladin interactive sky atlas \citep{2000A&AS..143...33B}. Aladin enables one to simultaneously visualize different wavelengths of images from the Panoramic Survey Telescope and Rapid Response System (PanSTARRS) PS1 \citep{2016arXiv161205560C} and the INT Photometric H-Alpha Survey (IPHAS) \citep{2005MNRAS.362..753D}.
Recently, images from the DECaPS2 survey \citep{2023ApJS..264...28S} have also been incorporated into the search.

The basic methodology is to look for compact features which show up in red photometric bands that cover the strongest optical lines from shock-excited emission (H$\alpha$ 656.3 nm, [{N\sc{ii}}] 654.8, 658.3 nm, [{S\sc{ii}}] 671.6, 673.1 nm, and [{O\sc{i}}] 630.0, 636.4 nm). Since this flux excess is absent in visual and infrared filters, it renders the detection of candidate HHOs feasible.

In addition, narrowband imaging was performed using the TAUKAM \cite{2016SPIE.9908E..4US} prime focus camera on the 2-m Alfred-Jensch telescope (AJT), which provides an aperture of 1.34\,m in the Schmidt configuration. TAUKAM features a 6144\,x\,6160 pixel e2v CCD, offering a 1.75\raisebox{.5ex}{$\scriptstyle{\square}$}\degr{} field-of-view (FoV) at a 0.775\arcsec{} pixel scale. Full frame images were obtained during 2024 fall and 2025 spring using H$\alpha$ and [{S\sc{ii}}] filters, as well as the $i$ filter of the SDSS filter set. Two images per band were taken to facilitate the removal of cosmics, with total exposure times of 1200\,s for the narrowband filters and 360\,s for the $i$ band. The world coordinate system has been established using the astrometry.net code \citep{2010AJ....139.1782L}.

Supplementary imaging from the Wide Field Infrared Survey Explorer (WISE) \cite{2010AJ....140.1868W} and IRAC@Spitzer \citep{2004ApJS..154...10F} has been used to check the presence of shocked emission, mainly from H$_2$ and/or CO in the W2 (4.6\,\micron) and I2 (4.5\,\micron) bands, as well as jet-driving infrared (IR) sources. For this purpose, the ``finderchart'' and ``viewer'' services provided by the NASA/IPAC Infrared Science Archive (IRSA) have been primarily used. In order to characterize the variability of the objects, (NEO)WISE photometry in the W1 (3.4\,\micron) and W2 bands was retrieved from IRSA. 

The complete sample of identified objects, sorted according to the right ascension, is listed in Tab.\ref{tab:coords} The top part is shown below to explain the nomenclature, where ``ID'' is the Bres identifier along with the running number. The plus sign in the ``T'' column marks objects that have been observed with TAUKAM. The plus sign in the next column indicates the possible HHO nature. 
The rightmost column provides SIMBAD identifiers if available. 
Where candidate HHOs were identified near a Bres object, their J2000 equatorial coordinates are listed below that object, with each candidate labeled by the Bres number plus a letter (e.g., 6A, 6B, ...).
\renewcommand{\tabcolsep}{3pt}
\begin{table}[h]
\centering
\captionsetup{labelformat=abbreviated}
\caption{}
\label{tab:coords}
\begin{tabular}{|c|m{1.5cm}|m{1.45cm}|m{.2cm}|c|l|}
\hline
{ID} & RA & Dec & T &HH&Simbad Identifier\\
 2 & 00 09 46.0 & +65 33 36 & &&\\              
 3 & 00 12 51.1 & +60 29 29 & &&\\             
 4 & 00 57 20.0 & +62 39 55 & &&ZOAG123.61-0.20 \\               
 5 & 01 20 03.0 & +66 51 14 & + & + &HH1227\\ 
 ... &&&&&\\
  7 & 03 54 37.2 & +53 12 36 & + & +&IRAS 03510+5301\\             
 \hphantom{0}7A  & 03 54 47.7 & +53 10 36 & + &+&\\
 ... &&&&&\\
\hline
\end{tabular}
\end{table}

\section{Characterization of individual objects}\label{sec4}


\subsection{Bres 1}
\begin{figure}[htbp]
\centerline{\includegraphics[width=0.75\linewidth]{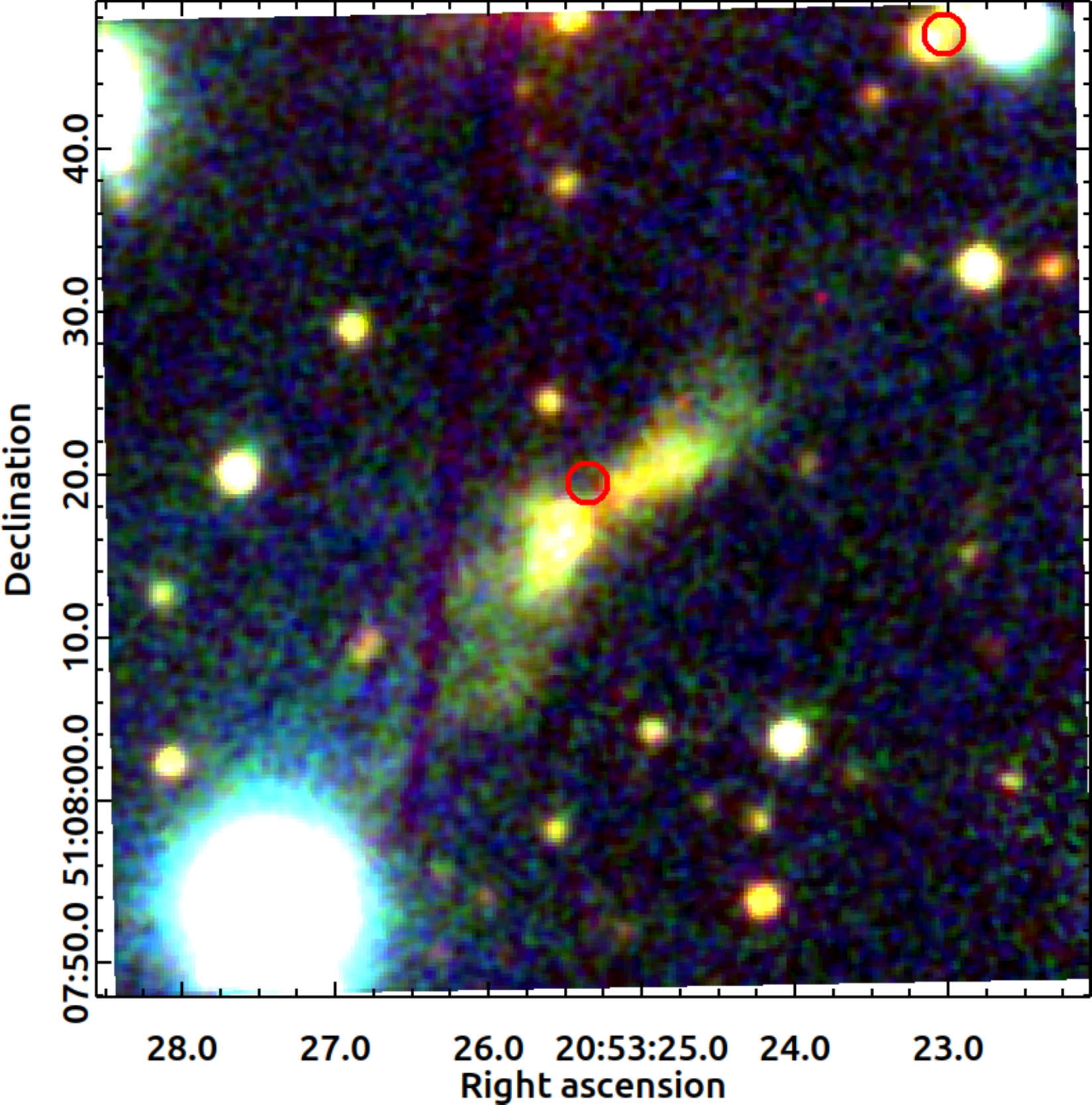}}
	\caption{PS1 $i,r$ and $g$ RGB composite for Bres~1. Red circles mark ALLWISE sources. }
    \label{fig1}
\end{figure}

We include this object (Fig.\,\ref{fig1}) described by \cite{2023RNAAS...7..254B} for the completeness. Its biconical nebula turned out to be a signpost of a YSO in an early stage of formation. Fig.\,\ref{fig1} shows an RGB color composite of PS1 $i, r$ and $g$ frames with a FoV of 1\raisebox{.5ex}{$\scriptstyle{\square}$}\arcmin{} that have been slightly smoothed by a 2D Gaussian to suppress noise. The green tint of the extended emission from the lobes points to scattered light from emission lines covered by the $r$ band, which is typical for HHOs.

The  slight displacement of the embedded YSO ALLWISE J205323.02+510846.9 with respect to the longitudinal axis is probably due to an extinction gradient in the foreground (cf. \citealt{2023RNAAS...7..254B}).

\subsection{Bres 2}
\begin{figure}[!htbp]
\centerline{\includegraphics[width=0.75\linewidth]{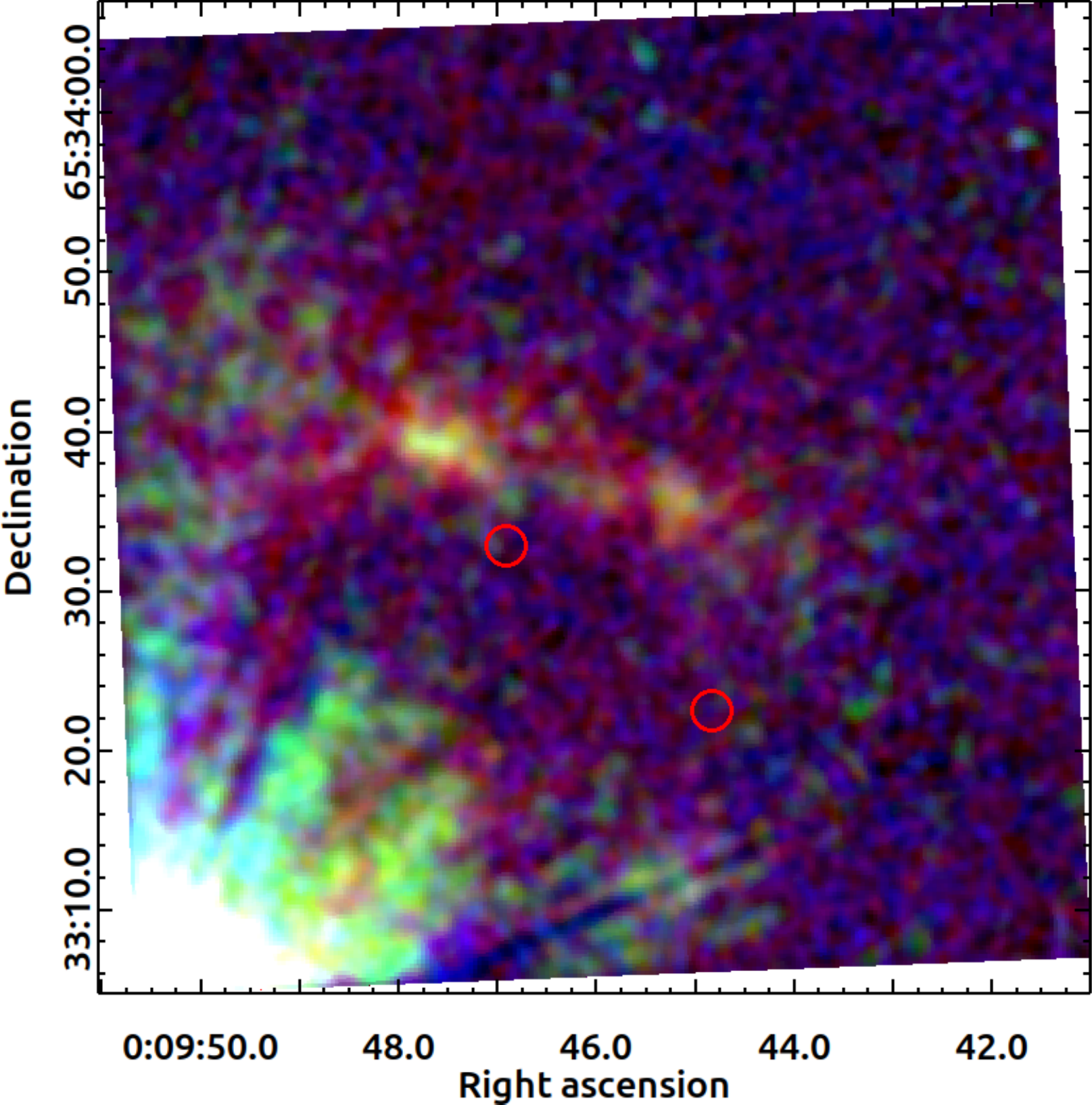}}
	\caption{Same as Fig.\,\ref{fig1} for Bres~2. }
    \label{fig2}
\end{figure}

 Bres\,2 (Fig.\,\ref{fig2}) is located in the dark cloud DOBASHI~3646 \citep{2011PASJ...63S...1D} which is associated with the Planck cold core PGCC G118.65+03.04 \citep{2016A&A...594A..28P}, a region of active star formation.
 Bres\,2 has a bipolar morphology suggestive of outflow lobes seen in scattered light. However, contrary to Bres\,1, the closest ALLWISE source is not located between the presumed lobes. This argues against the outflow interpretation. Thus, it is likely a reflection nebulosity of a different nature.

\subsection{Bres 3}
\begin{figure}[!htbp]
\centerline{\includegraphics[width=0.75\linewidth]{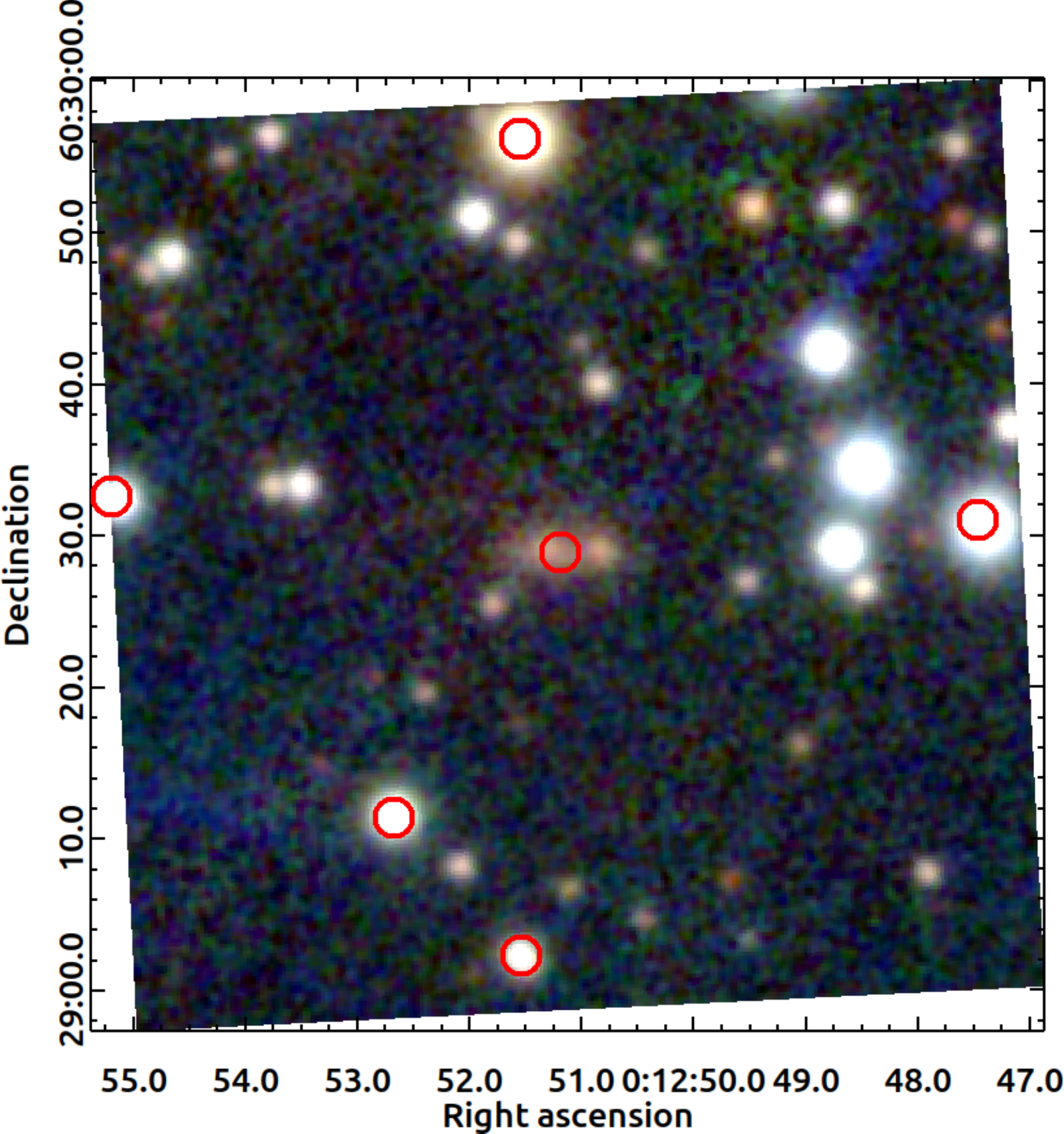}}
	\caption{Same as Fig.\,\ref{fig1} for Bres~3.}
    \label{fig3}
\end{figure}

 Bres 3 (Fig.\,\ref{fig3}) has a double structure and is associated with the ALLWISE source J001251.18+602928.9. Its W1$-$W2 color index of 0.45\,mag does not point to an embedded YSO. IPHAS measurements \citep{2005MNRAS.362..753D} did not observe H$\alpha$ emission.  Furthermore, the lack of variability in the (NEO)WISE light curves and the non-detection by AKARI \citep{2015PASJ...67...50D} at far-infrared (FIR) wavelengths also indicate that this is not a YSO. 
 A recent study classifies this object as a galaxy \citep{2026A&A...706A.284T}.

\subsection{Bres 4}
\begin{figure}[!htbp]
\centerline{\includegraphics[width=0.75\linewidth]{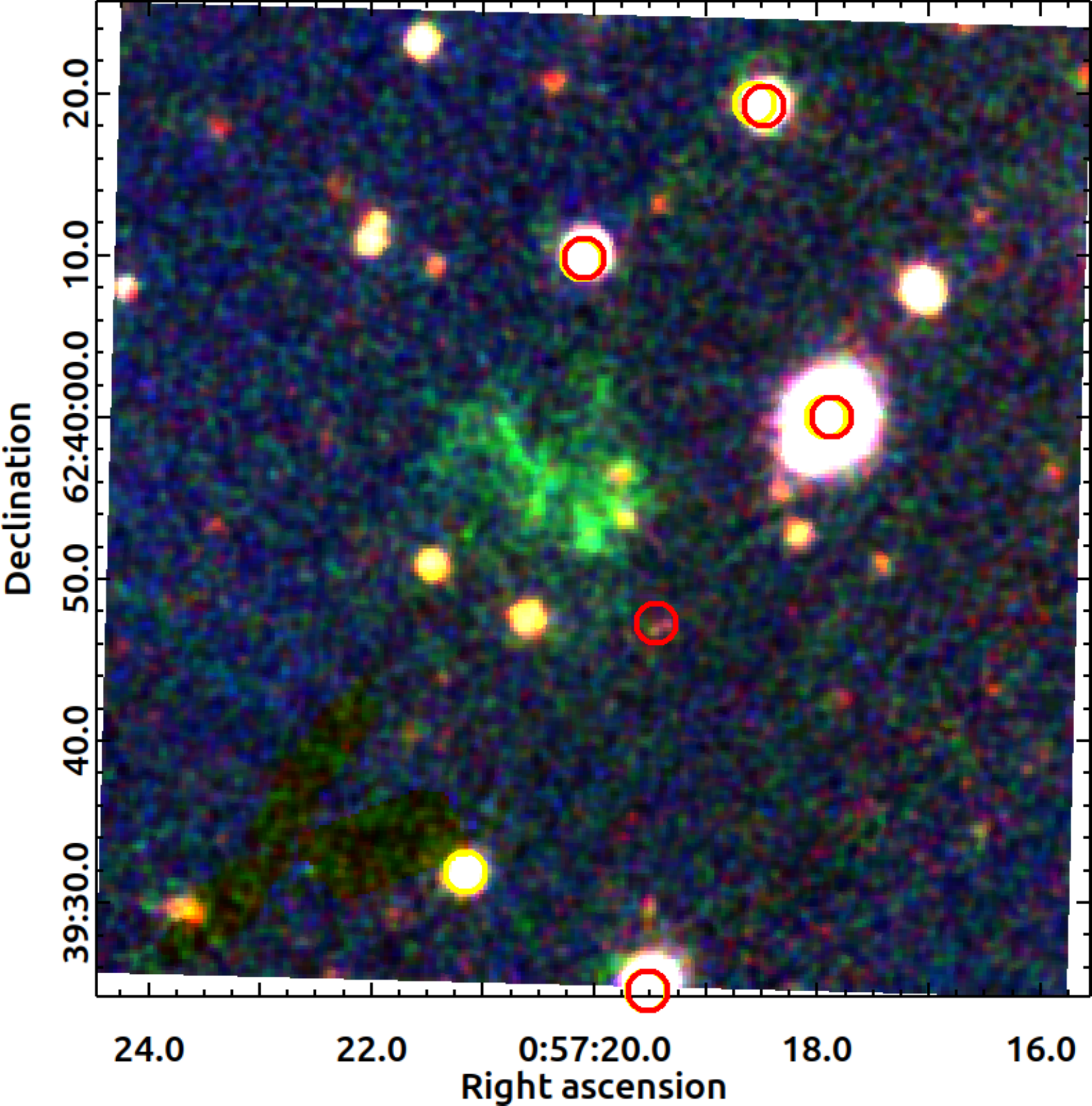}}
	\caption{Same as Fig.\,\ref{fig1} for Bres~4.}
    \label{fig4}
\end{figure}

This object  (Fig.\,\ref{fig4}) shows an irregular and patchy structure, dominated by H$\alpha$ emission as observed by IPHAS \citep{2005MNRAS.362..753D}. It is not associated with an infrared source, which rule out a YSO or PPN nature. Instead, it was classified as galaxy ZOAG\,123.61-0.20 \citep{1999A&AS..137..293W}.

\subsection{Bres 5}
\begin{figure}[!htbp]
\centerline{\includegraphics[width=0.75\linewidth]{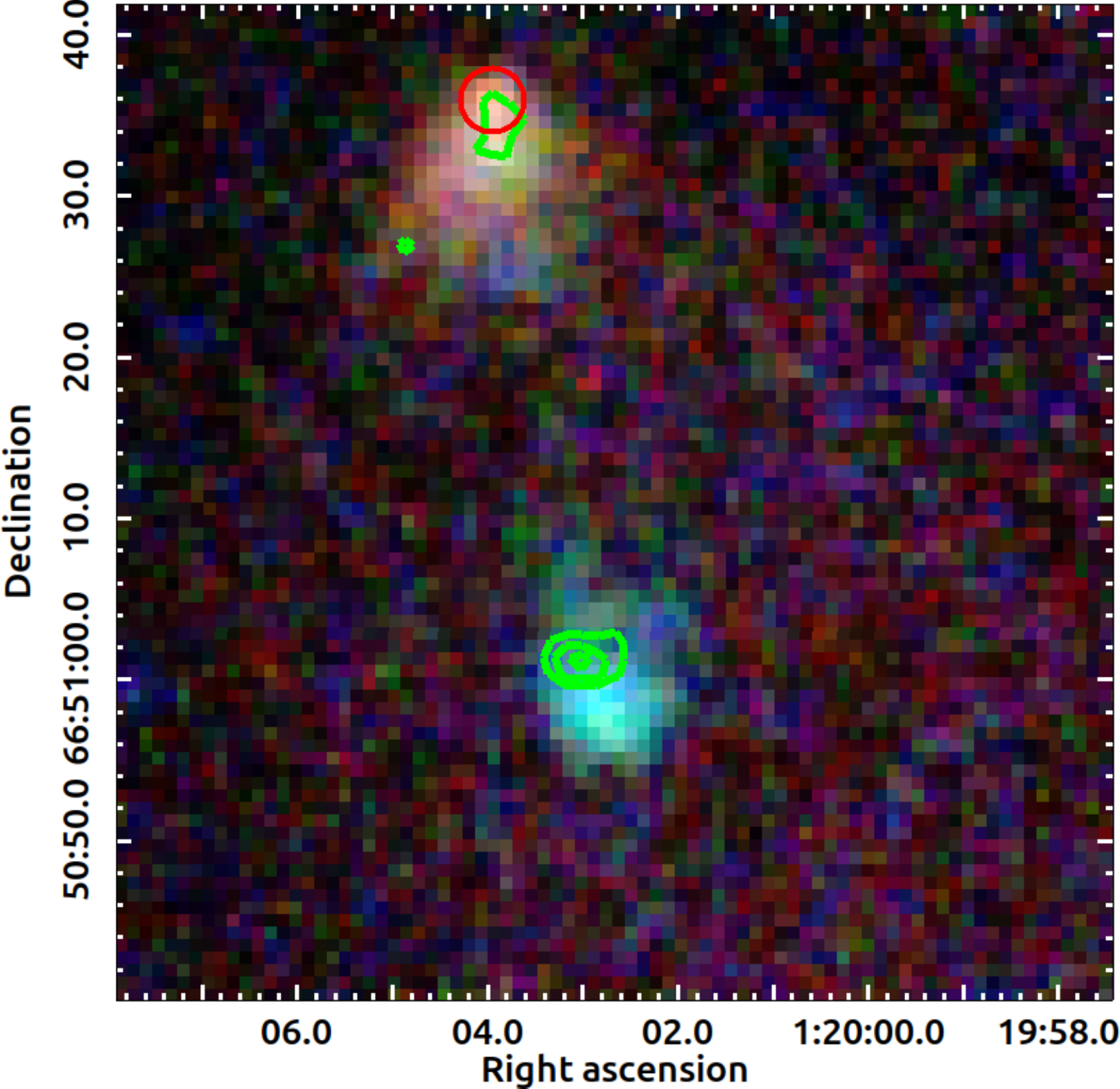}}
	\caption{TAUKAM $i$, H$\alpha$, and [{S\sc{ii}}] RGB composite of Bres~5.  The red circle marks the ALLWISE source while contours denote the emission in the R-band of POSS-II \citep{1991PASP..103..661R}.}
    \label{fig5}
\end{figure}

Fig.\,\ref{fig5} shows a $i$, H$\alpha$, and [{S\sc{ii}}] RGB composite, based on TAUKAM images. The HHO, which has been recently detected by another survey \citep{2024MNRAS.530.2068M}, can be easily recognized by its characteristic color. Its morphology resembles that of a bow shock of a protostellar jet. The cometary nebula north of it represents the optical signpost of the driving source IRAS 01166+6635, also known as ALLWISE J012003.93+665135.9. It is the only known YSO present in the dark cloud DOBASHI 3782 \citep{2011PASJ...63S...1D}.

Comparison of our narrowband frames with the R-band POSS-II image (green contours, \citealt{1991PASP..103..661R}) revealed the proper motion of the HHO. The angular speed, derived from the shift of 3\farcs6 and an epoch difference of 23 years, is 0\farcs156\,$\rm yr^{-1}$. With a distance of the bow shock from the YSO of 39\arcsec{} in November 2024, its dynamic age is about 250 years, assuming constant velocity. Taking into account the distance of the IRAS source of 250\,pc \citep{1989A&AS...80..149W}, it translates into a tangential velocity of almost 200\,km\,s$^{-1}$. This suggests that the YSO experienced an accretion burst in the recent past, which triggered strong jet activity. The cometary morphology of the associated reflection nebula indicates an intermediate inclination with respect to the line of sight, as suggested by radiative transfer models of YSOs, e.g., \cite{2003ApJ...591.1049W}. This would also explain the lack of the red-shifted outflow lobe and the counterjet, which are probably shadowed by both, circumstellar disk and dark cloud. The source is clearly variable, as indicated by the joint Stetson index \citep{1996PASP..108..851S} of the W1 and W2 magnitudes of $\sim$5.

\subsection{Bres 6}
\begin{figure}[!htbp]
\centerline{\includegraphics[width=.75\linewidth]{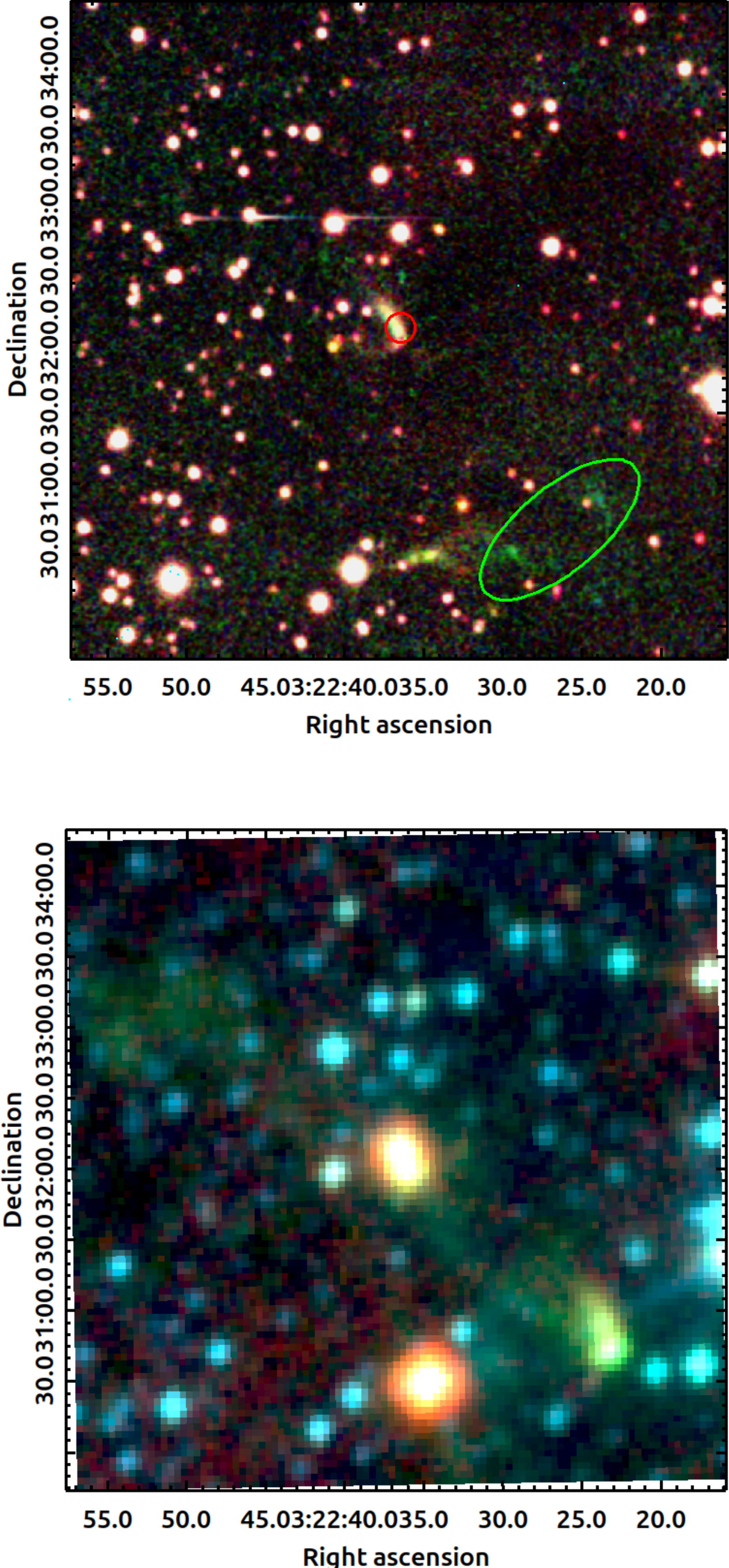}}
	\caption{{\bf Top: }TAUKAM $i$, H$\alpha$, and [{S\sc{ii}}] color composite for Bres~6 (center). The ALLWISE position is marked by the red circle, and the green ellipse encompasses the optical emission line features. {\bf Bottom:} RGB composite based on unWise \citep{2021RNAAS...5..168M} W3 (12\,\micron), W2, and W1 images. 
    }
    \label{fig6}
\end{figure}

The cometary-shaped optical nebulosity of Bres\,6 (Fig.\,\ref{fig6}, top) is associated with a close pair of objects listed as 2MASX\,03223621+6332054 in the 2MASS Extended Catalog \citep{2006AJ....131.1163S} that corresponds to the ALLWISE source J032236.42+633206.2 (Fig.\,\ref{fig6}, bottom). Although emission in H$\alpha$, and [{S\sc{ii}}] can barely be seen at the source, there is a small knot north of it and an arcuate feature south-west. The W2 (4.6\,\micron) image, covering shock-excited emission lines from H$_2$ and possibly the fundamental transition of CO, 
shows a splendid bipolar outflow. While the line emission is more widespread in the north-east, there is an extremely bright shock in the south-west. It is also seen in the 2MASS K-band image, and a glimpse of it is obvious in the optical narrowband filters. The NEO(WISE) light curve hints at a periodicity of about six years, which is accompanied by a shift of the centroid position. This points to the presence of a binary system. Variability-induced image motion allows for recovery of individual light curves of the two components \citep{Stecklum_2025}.

\subsection{Bres 7}
\begin{figure}[!htbp]
\centerline{\includegraphics[width=.775\linewidth]{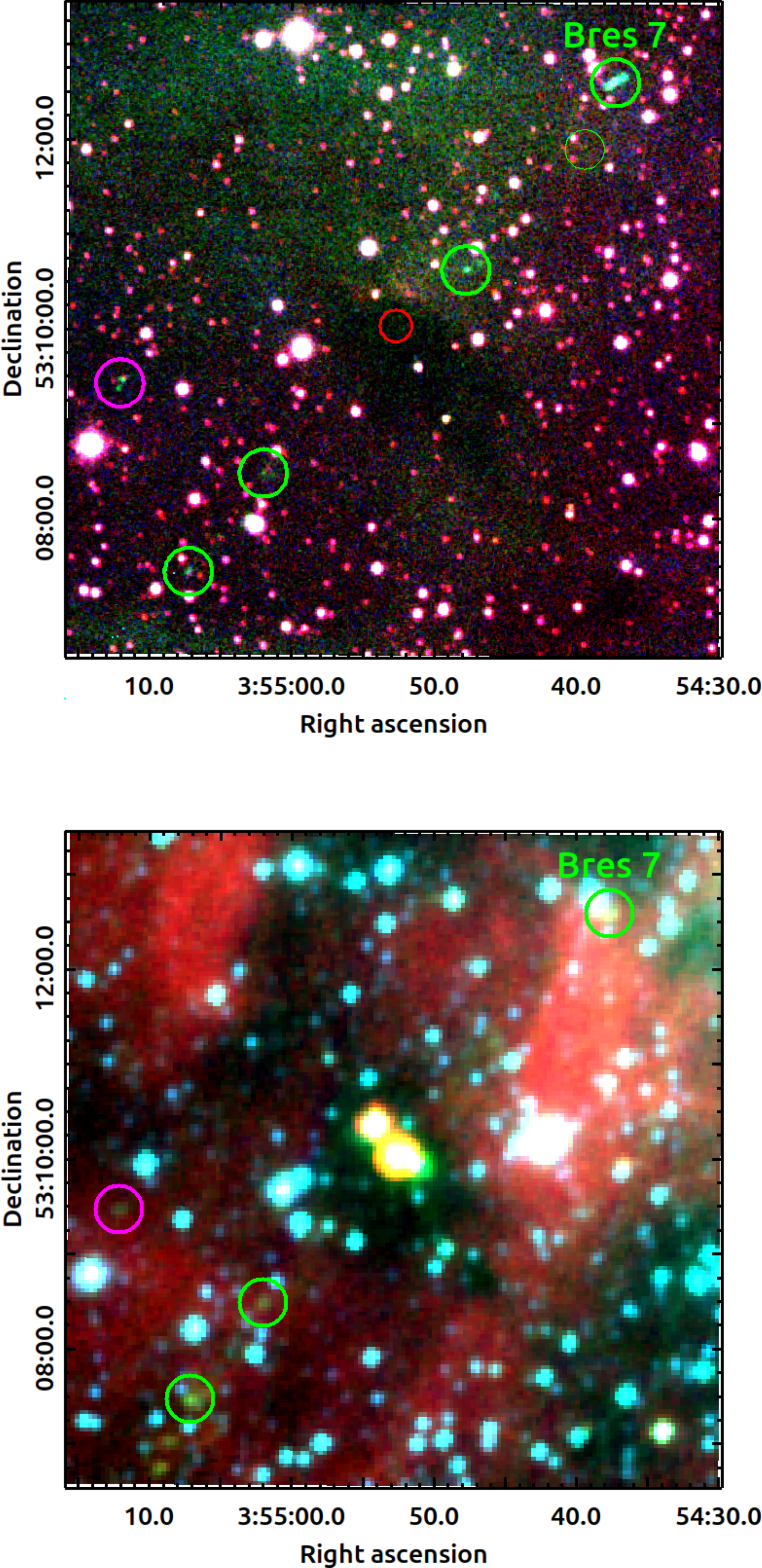}}
    \caption{Same as Fig.\,\ref{fig6} for Bres~7.  Green circles mark jet knots. Bres\,7 is the jetlike feature to the north-west. The emission knot indicated by the magenta circle likely belongs to a different jet.
    }
    \label{fig7}
\end{figure}

Bres\,7 appears to be a bright HHO with a pronounced linear morphology  (Fig.\,\ref{fig7}, top), but isolated, since there is no YSO in its immediate surroundings. 
It can also be seen in the POSS-II red image \citep{1991PASP..103..661R}. There is no noticeable proper motion within the epoch difference of 34 years between the POSS-II and TAUKAM imaging.

Closer inspection of the TAUKAM images revealed fainter emission-line knots which line-up to the south-east and run across an anonymous small dark cloud. The unWISE color image (Fig.\,\ref{fig7}, bottom) shows the presence of YSOs and HHO counterparts seen in molecular emission of H$_{\rm 2}$ and possibly CO. The dark cloud contains IRAS 03510+5301 which corresponds to ALLWISE J035452.63+531001.6. This source consists of two components that drive orthogonal outflows, as traced by shock-excited  H$_{\rm 2}$ emission at 2.122\,\micron{} \citep{2016MNRAS.462.1444F}. The one identified by us in the optical and at mid-infrared wavelengths corresponds to MHO1371. Both components of the ALLWISE source show extraordinary variability.

\subsection{Bres 8}
\begin{figure}[!htbp]
\centerline{\includegraphics[width=.75\linewidth]{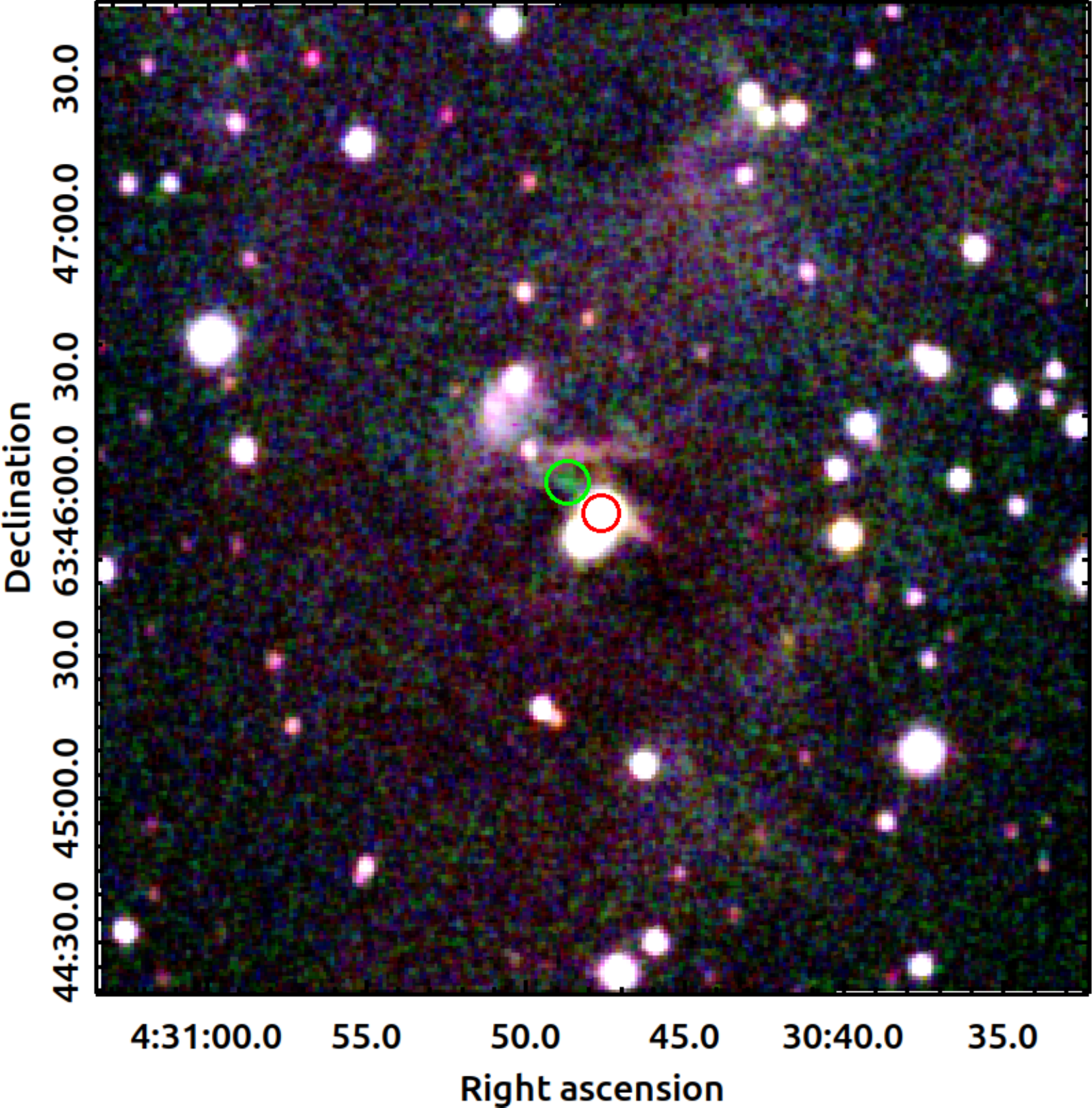}}
    \caption{Same as Fig.\,\ref{fig5} for Bres~8. The green circle marks the one-sided jetlike feature.
    }
    \label{fig8}
\end{figure}

A visual binary is located in the center of Fig.\,\ref{fig8}. Closer inspection reveals that the north-east component, which corresponds to ALLWISE J043047.58+634559.4 aka IRAS 04261+6339, shows a short linear emission-line feature toward the north-west that resembles a micro-jet. Remarkably, opposite to the jetlike feature, a similar one is seen in the red. It might represent the reddened counter-jet, perhaps mainly emitting in H$_{\rm 2}$.  A previous study of the IRAS source by \cite{2004AJ....128..375M} claimed that both stars represent YSOs, surrounded by a huge disk which is barely seen in scattered light. This hypothesis has to be withdrawn given the GAIA photogeometric distances, which amount to 2.5\,kpc for the IRAS source and 0.93\,kpc for the star to the south-west \citep{bailer-jonesEstimatingDistancesParallaxes2021}. The optical lightcurve of the IRAS source suggested that it is a long period variable \citep{2018yCat.1345....0G}.

\subsection{Bres 9}
This object is identical with HH378A \citep{1997IAUS..182P..91D} and has been kept in the list for completeness.

\subsection{Bres 10}
\begin{figure}[!htbp]
\centerline{\includegraphics[width=.75\linewidth]{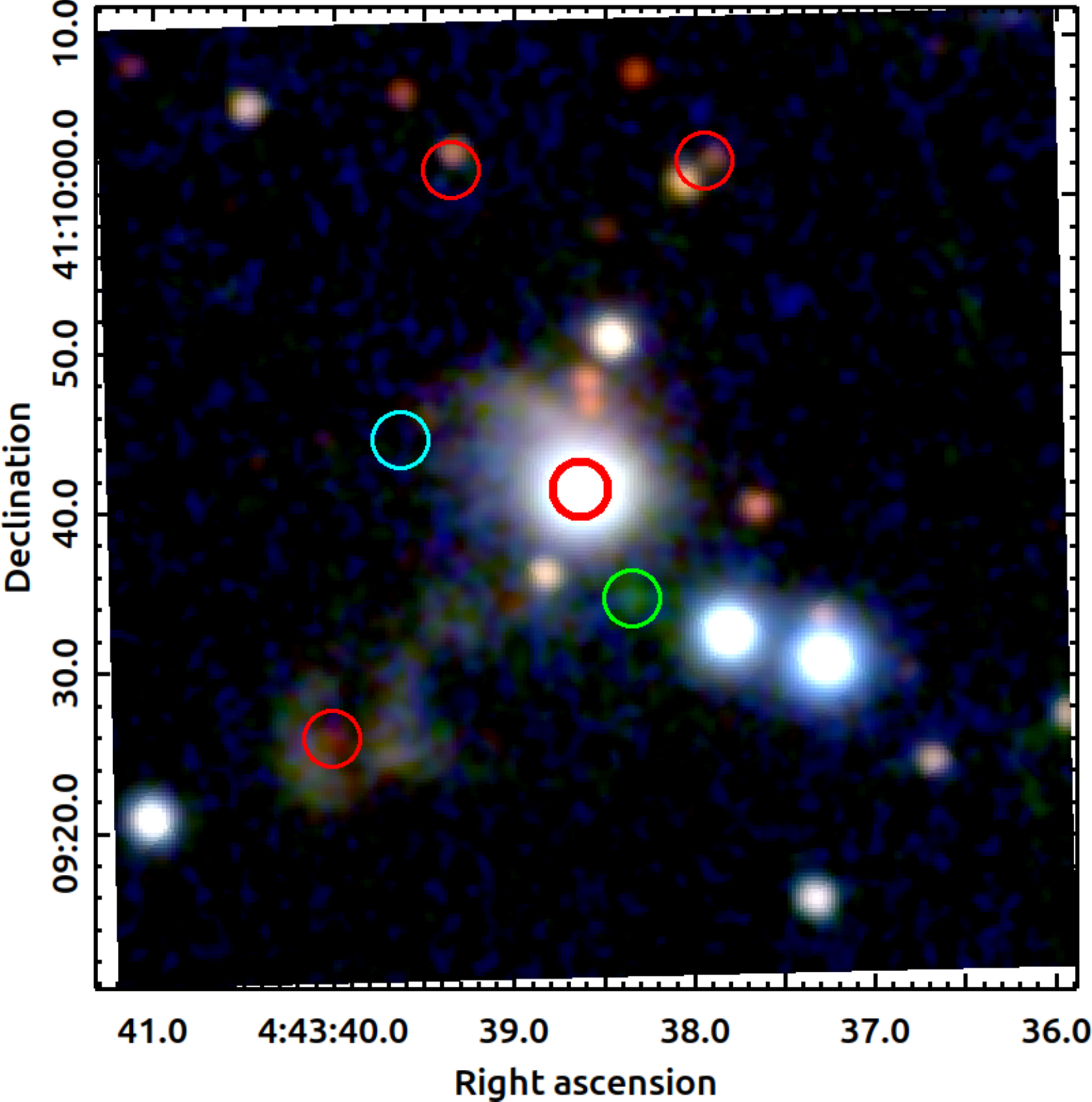}}
    \caption{Same as Fig.\,\ref{fig1} for Bres~10. The cyan circle marks the position of the Planck cold core, while the green one denotes the HHO. The south-eastern ALLWISE source might drive a jet which gives rise to the HHO.
    }
    \label{fig10}
\end{figure}

Bres\,10 (Fig.\,\ref{fig10}) is the faint reflection nebula north-east of the optical star UCAC2~46104304, classified by Simbad as PN candidate (\citealt{2009A&A...502..113V}) at the center. However, the presence of the nearby Planck cold core PCCS2E 217 G162.70-03.11  \citep{2016A&A...594A..28P} challenges this classification. Moreover, there are several reddened sources in the field, as well as nebulosity to the south-east, which point to a small star forming region. Meanwhile, UCAC2~46104304 has been recognized as a candidate Herbig Ae/Be star \citep{2020A&A...638A..21V}.
About 7\arcsec{} south of it, there is compact emission in the PS1 $r$ image that traces a HH candidate which has been confirmed by TAUKAM imaging. Its alignment with the orientation of the biconical nebulosity surrounding the ALLWISE source J044340.00+410925.8 to the south-east suggests the presence of an outflow from this YSO. It is located at the peak position of the FIR emission observed by AKARI \citep{2015PASJ...67...50D}.

\subsection{Bres 11 and Bres 12}
\begin{figure}[!htbp]
\centerline{\includegraphics[width=.75\linewidth]{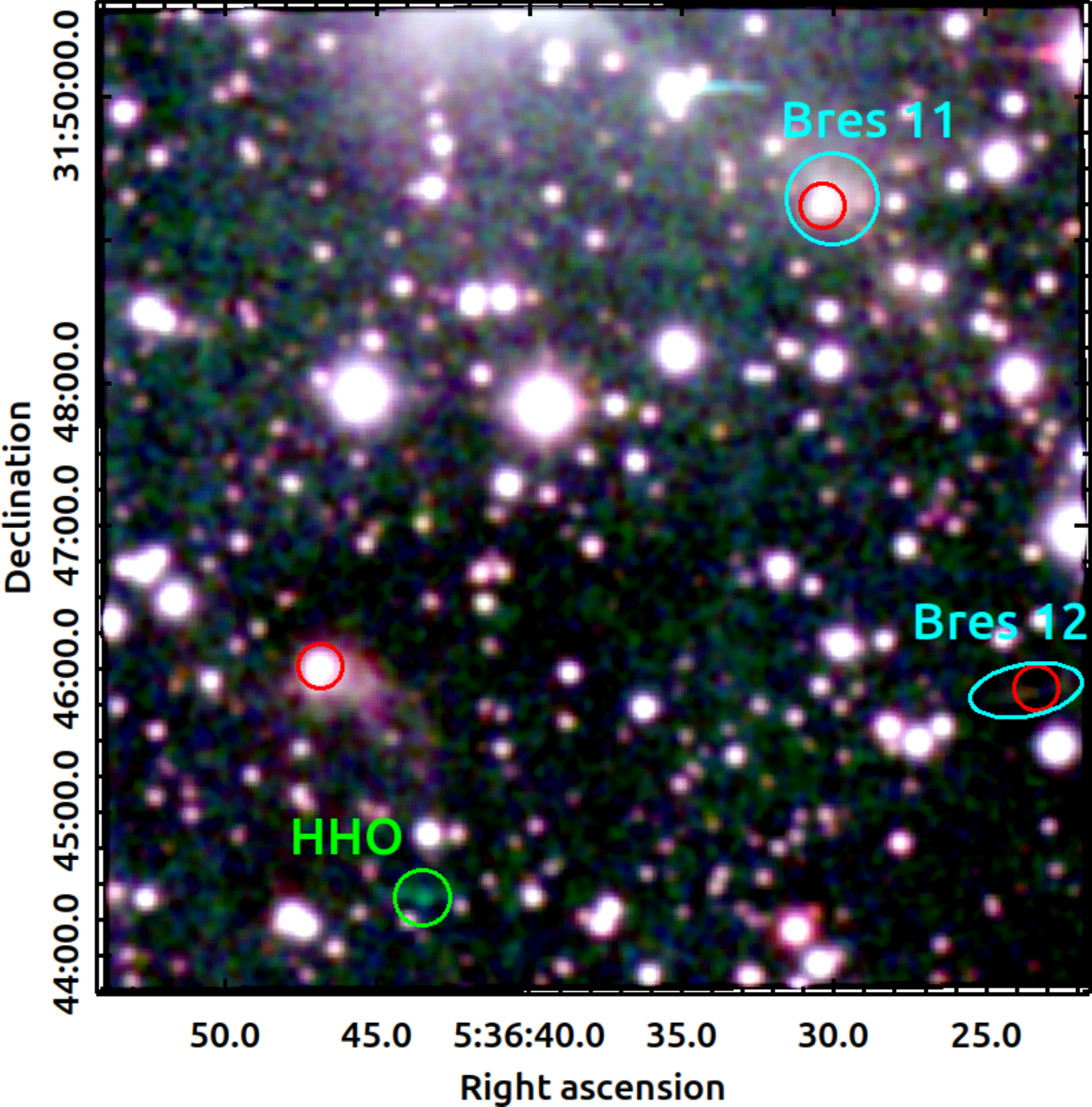}}
    \caption{Same as Fig.\,\ref{fig5} for Bres~11 and 12. The sources and the HHO are labeled.
    }
    \label{fig11}
\end{figure}

The small reflection nebulosity labeled Bres 11 in Fig.\,\ref{fig11} is north-east of the illuminating star, ALLWISE J053630.36+314914.3. It has red colors and features H$\alpha$ emission as indicated by the (H$\alpha-r$) excess of $-0.69$\,mag  \citep{2020A&A...638A..18M} which supports its YSO nature. The median photogeometric distance to Bres 11 is 1\,kpc \citep{bailer-jonesEstimatingDistancesParallaxes2021}.

Bres~12 is a very faint elongated bipolar feature of $\sim$20\arcsec{} length that almost runs east-west. It is at the detection limits of the TAUKAM imaging and the surveys. At the apex of the brighter western lobe there is ALLWISE J053623.31+314551.8 which features an SED typical for embedded YSOs. It is associated with the Herschel/PACS Point Source HPPSC160A\_J053623.3+314552 \citep{2017arXiv170505693M}. Thus, the eastward conical nebula probably represents the cavity created by the blue-shifted outflow. 4\farcm{}5 south-east of it, a compact candidate HHO has been found in the TAUKAM narrowband images. Since the connecting line to Bres~12 has about the same orientation as the blue-shifted lobe, it might be excited by this flow component. However, 1\farcm{}7 north-east of the HHO is another YSO, ALLWISE J053646.84+314601.1 aka IRAS~05335+3144. Its cometary appearance is also aligned with the line connecting with the HHO. Thus, it could also be the driving source. Proper motion measurements of the HHO are required to identify the driving source.

\subsection{Bres 13 and 14}
\begin{figure}[!htbp]
\centerline{\includegraphics[width=.775\linewidth]{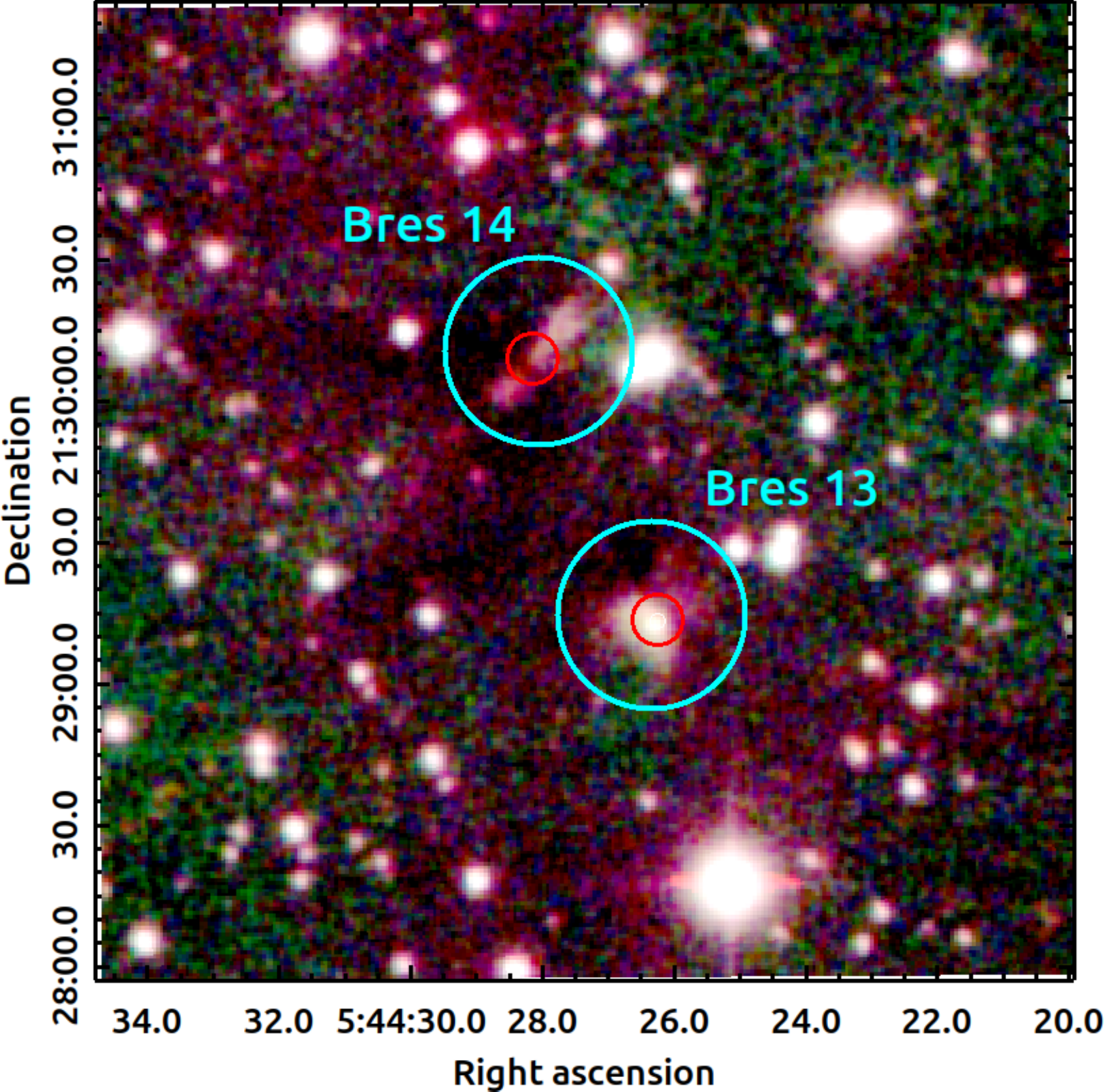}}
    \caption{Same as Fig.\,\ref{fig5} for Bres~13 and 14.
    }
    \label{fig13}
\end{figure}

These two objects (Fig.\,\ref{fig13}) are situated in an elongated dark cloud that hosts the Planck cold core PGCC G186.21$-$04.08 \citep{2016A&A...594A..28P}. The southern one, Bres 13, has a biconical shape that resembles outflow cavities with a substantial opening angle. GAIA detected it, which allowed us to estimate its median photogeometric distance of 600\,pc \citep{bailer-jonesEstimatingDistancesParallaxes2021}. Its YSO nature was already recognized by \cite{2016MNRAS.458.3479M} and \cite{2020A&A...638A..21V}. The weakness of its north-west lobe points to a relatively small inclination toward the line of sight.

The northern Bres 14 has a similar morphology, but with a much smaller opening angle. This points to an earlier evolutionary stage compared to its companion Bres 14. Although the position angles of the symmetry axes of bipolar nebulae seem to be similar, the space orientation is different since the brighter north-west lobe of Bres 14 is facing toward us, while for Bres 13 this holds for the south-east lobe. 

Both nebulosities are associated with ALLWISE sources, which are located in dark lanes between the cavities. Their (NEO)WISE photometry shows variability with a possible periodicity of the order of $\sim$5 yr. Despite their YSO nature, no HH-like emission features have been found for both sources. This might hint at recent periods of low accretion activity.

\subsection{Bres 15}
\begin{figure}[!htbp]
\centerline{\includegraphics[width=.75\linewidth]{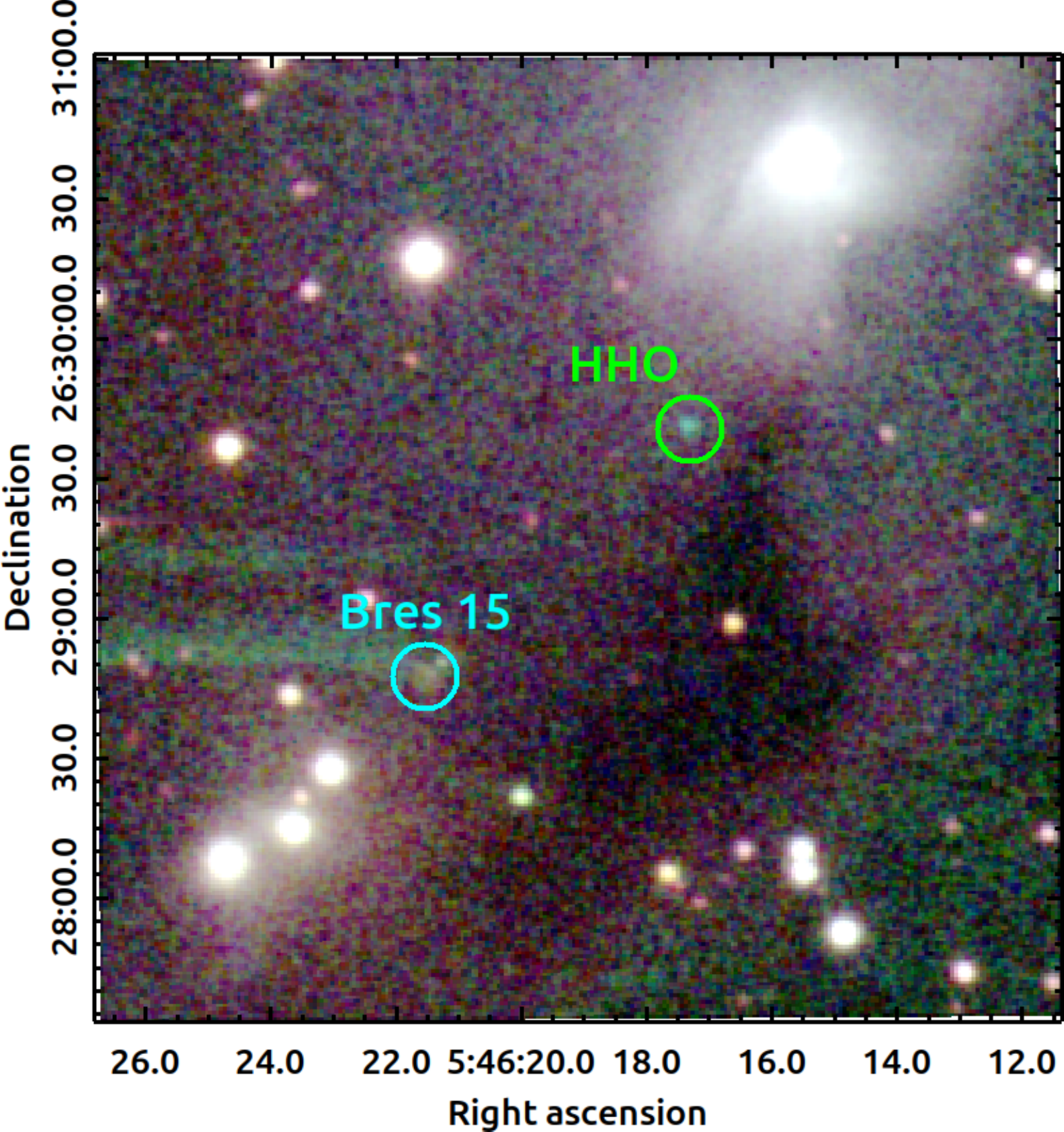}}
    \caption{Same as Fig.\,\ref{fig6} for Bres~15. The object as well as the newly found HHO are marked by the blue and green circles. The green streak at the eastern border is an image artifact.
    }
    \label{fig15}
\end{figure}

This object represents a small nebulous patch next to an anonymous dark cloud, with the reflection nebula GN 05.43.3 at its north-eastern border. Although Bres 15 does not show an emission line excess, $\sim$80\arcsec{} north-west of it, a new HHO has been found (Fig.,\ref{fig15}). When searching for the driving source of a jet which could give rise to this HHO, no obvious candidate could be identified. The far-infrared emission associated with the star TYC 1870-277-1 which illuminates the reflection nebula originates from dust grains in its outer regions. GN 05.43.3 is probably located on the outskirts of the dark cloud. If so, the median photogeometric distance of 750\,pc for TYC 1870-277-1 holds for the entire complex.


\subsection{Bres 16}
\begin{figure}[!htbp]
\centerline{\includegraphics[width=.75\linewidth]{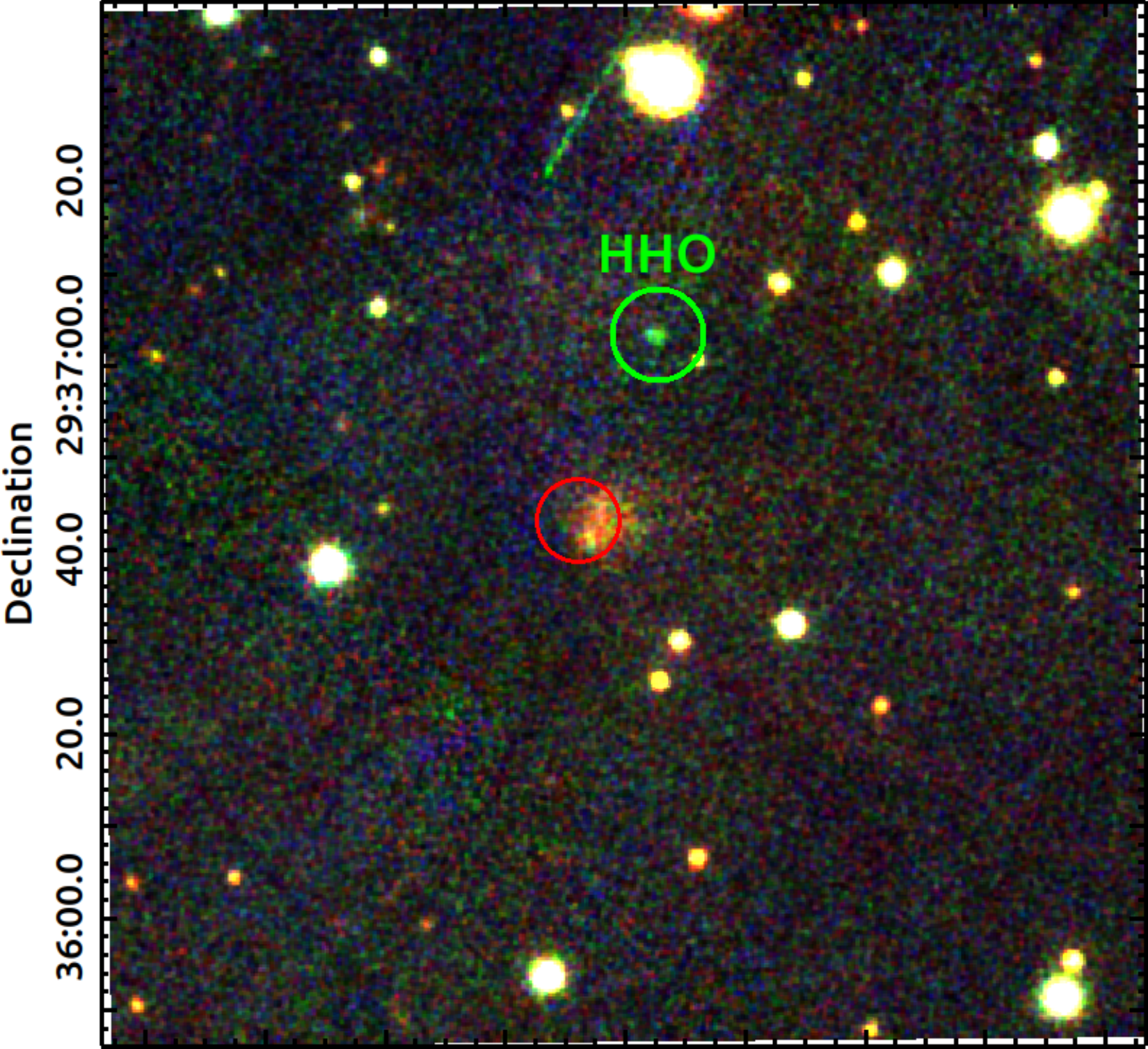}}
    \caption{Same as Fig.\,\ref{fig6} for Bres~16 (center). The newly found HHO is marked by the green circle. 
    }
    \label{fig16}
\end{figure}

The PS1 images show that Bres 16 (Fig.\ref{fig16}) has a conical morphology with an orientation toward the north-west. It is listed in the 2MASS All-Sky Extended Source Catalog \citep{2006AJ....131.1163S} as 2MASX~06084229+2936428 and associated with ALLWISE J060842.38+293643.1. About 20\arcsec{} north of its apex, a compact feature is present in the $r$ image, which is missing in the other bands. It is also present in the POSS-II red image \citep{1991PASP..103..661R} without noticeable proper motion between the two epochs. Thus, we identify it as a HHO candidate. This provides evidence for the YSO nature of Bres 16, with the conical nebula likely representing the blue-shifted outflow lobe. The invisibility of the red-shifted lobe suggests an intermediate inclination with respect to the plane of the sky. It seems to have been mistakenly identified as a galaxy \citep{2025arXiv250813267T}.

Remarkably, the (NEO)WISE photometry shows that the mid-IR brightness of the  IR source faded by about 2\,mag in W1 and by 1.5 mag in W2 over nearly 15 years. The fluxes from the AKARI/IRC mid-IR all-sky survey \citep{2010A&A...514A...1I}, carried out three years before the first WISE epoch, are consistent with this trend. It seems plausible that the YSO has experienced an accretion burst before 2007 and is on the way back to quiescence.

\subsection{Bres 17}
\begin{figure}[!htbp]
\centerline{\includegraphics[width=.775\linewidth]{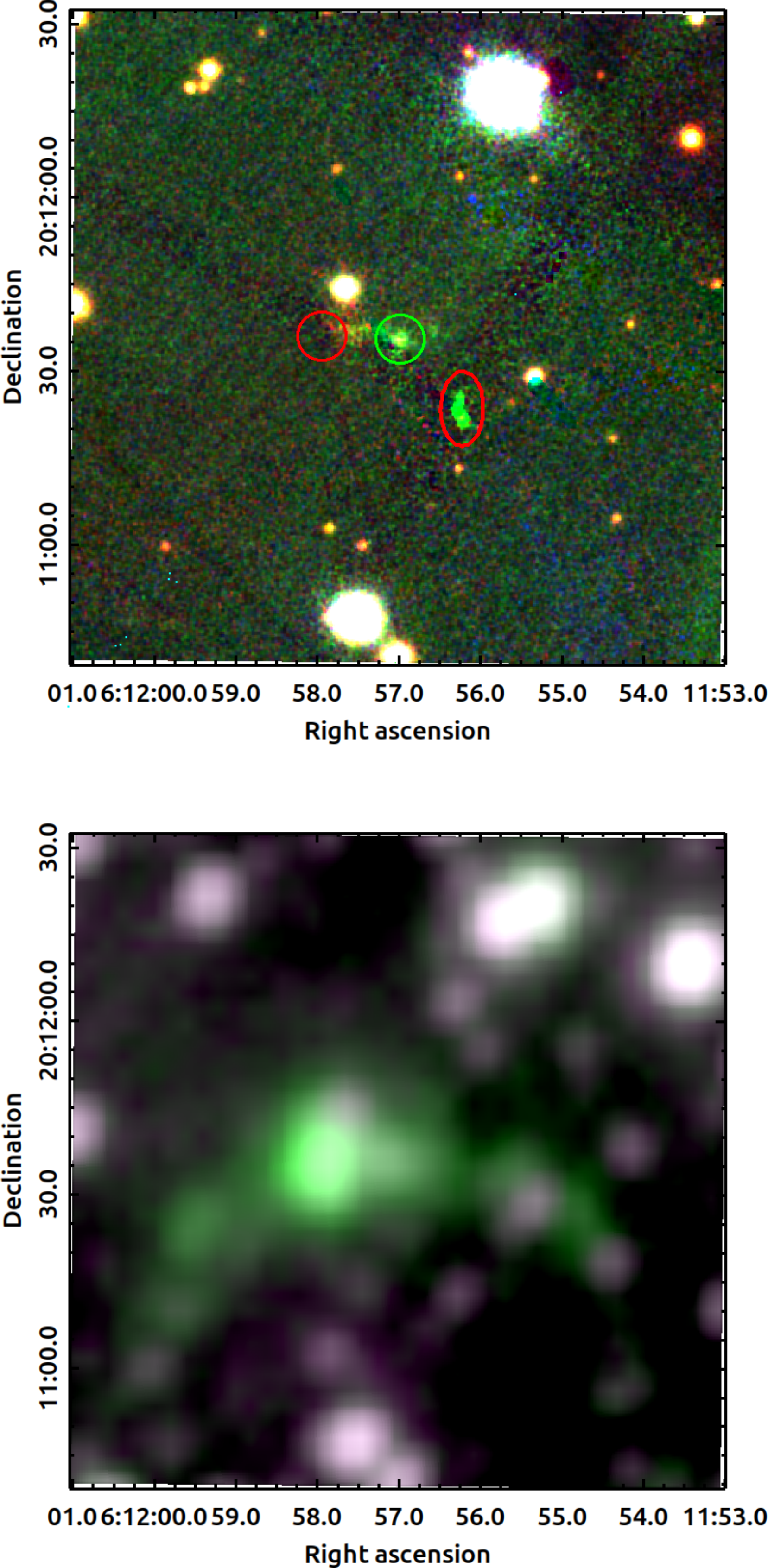}}
    \caption{{\bf Top:} Same as Fig.\,\ref{fig1} for Bres~17. The HHO and the ALLWISE source are marked by circles, while the red ellipse denotes an image artifact. {\bf Bottom:} unWISE RGB image based on W3, W2, and W1 frames.
    }
    \label{fig17}
\end{figure}

This object is marginally resolved in the PS1 images (Fig.\,\ref{fig17}, top) and is listed in the IGAPS point source catalog \citep{2020A&A...638A..18M} as a strong H$\alpha$ emitter with a faint continuum. This supports classifying it as HHO. It is very likely part of a highly bent outflow as indicated by the unWISE color composite  (Fig.\,\ref{fig17}, bottom), driven by the YSO ALLWISE J061157.93+201135.8 east of the HHO. The (NEO)WISE photometry suggests a periodicity of \,$\sim$10\,yr.


\subsection{Bres 18}
\begin{figure}[!htbp]
\centerline{\includegraphics[width=.75\linewidth]{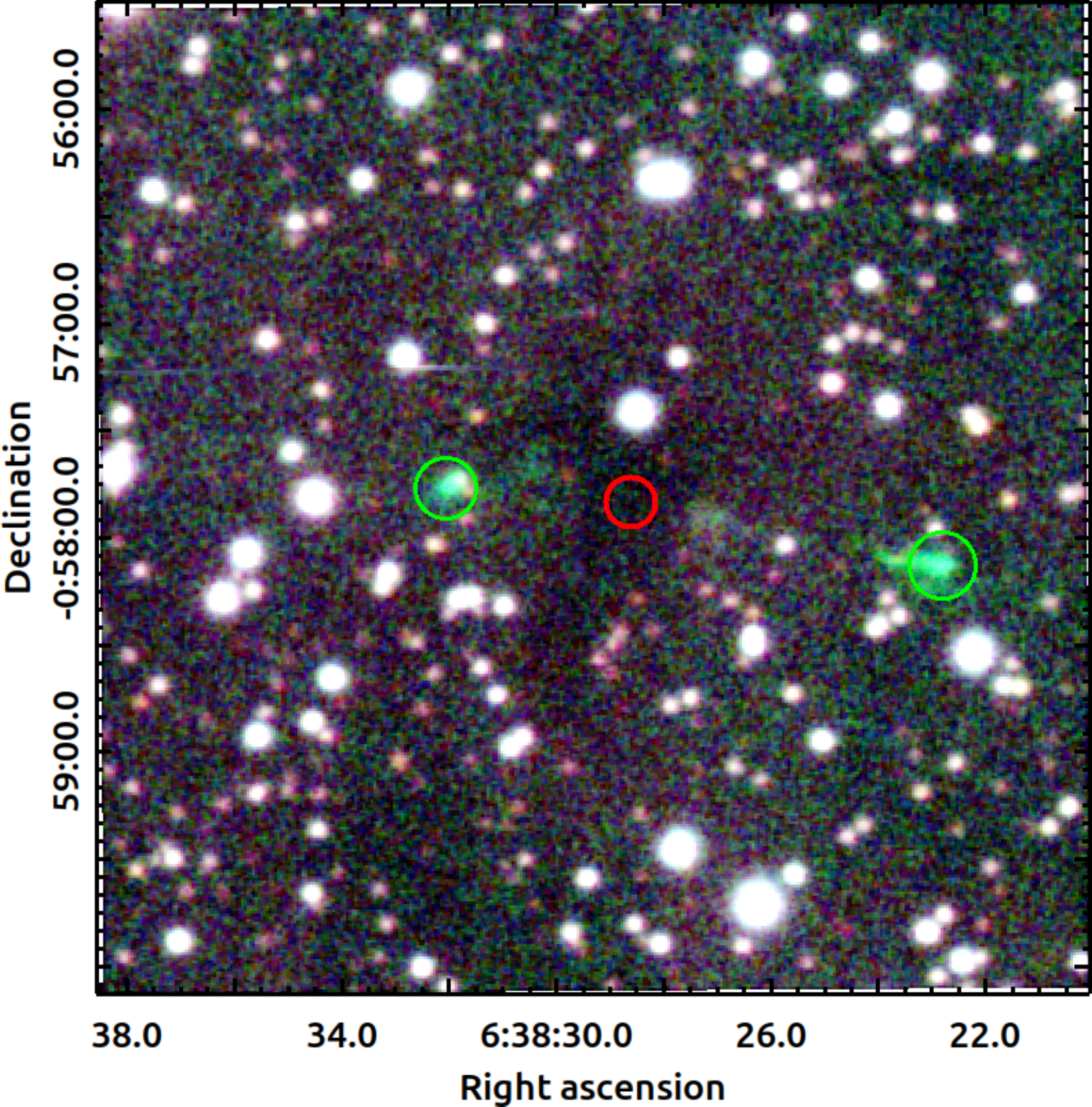}}
    \caption{Same as Fig.\,\ref{fig5} for Bres~18. The red circle marks the position of the AllWISE source, which lies close to the line connecting the jet bow shocks (green circles). 
    }
    \label{fig18}
\end{figure}

Bres 18 (Fig.\,\ref{fig18}) coincides with a dark cloud which corresponds to the Planck cold core PGCC G212.33-03.33 \citep{2016A&A...594A..28P}. It hosts the IRAS source IRAS~06359-0055. The CO(J=1$-$0) emission line associated with the IRAS source was measured by \cite{2002ApJS..141..157Y} who derived a kinematic distance of 1.08\,kpc. Wings in the line have been observed which might point to a molecular outflow. The TAUKAM narrowband imaging revealed HHOs on either side of the dark cloud, which points to the presence of a bipolar jet in the east-west direction. No obvious proper motion of the bow shocks could be detected within the epochs of the POSS-I (1955-11-18), POSS-II (1997-02-08) and TAUKAM (2024-12-29) imaging.

Among the AllWISE sources in the dark cloud area, J063828.58-005750.0 is closest to the line connecting the HHOs. Its W3$-$W4 color index of almost 5\,mag indicates that it is deeply embedded.
The (NEO)WISE photometry points at a periodicity of 7.4\,yr. Small shifts of the photocenter with the same period in north-south directions have been found, which seems to indicate the presence of a nearby companion.

\subsection{Bres 19}
This object has a particularly spatial appearance. It is located in an anonymous dark cloud that hosts IRAS~08244-4100 which Simbad lists as a mid-infrared source. Its morphology is cometary-shaped and oriented north-south. At the northern tip is the extended 2MASS source 2MASX~08261238-4110295, which is the near-infrared counterpart of ALLWISE J082612.36-411029.1. The 2MASS K and IRAC I2 (4.5\,\micron) images show that it is a double source. No evidence was found for an HH candidate. The shape of the reflection nebula might result from the one-sided illumination of the surface of the dark cloud, which is elongated toward the south.

It shows a brightness fading in the (NEO)WISE photometry similar to Bres~16, although not as strong. Nevertheless, it might be related to a post-burst behavior as well.
\begin{figure}[!htbp]
\centerline{\includegraphics[width=.75\linewidth]{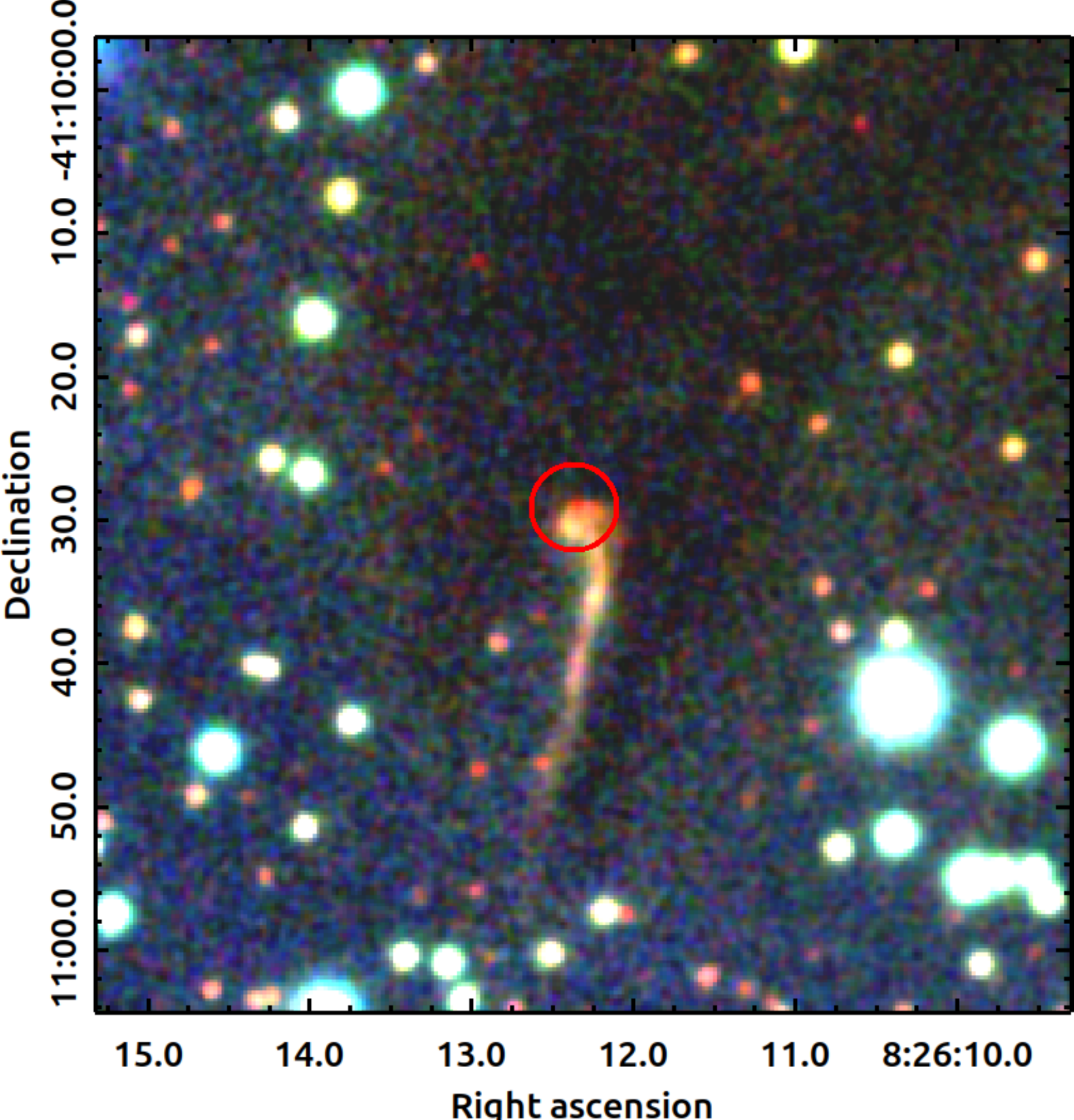}}
    \caption{RGB image based on DECaPS2 $i$, $r$, and $g$ frames for Bres~19. The red circle marks the position of the AllWISE source.
    }
    \label{fig19}
\end{figure}

\subsection{Bres 20}
\begin{figure}[!htbp]
\centerline{\includegraphics[width=.75\linewidth]{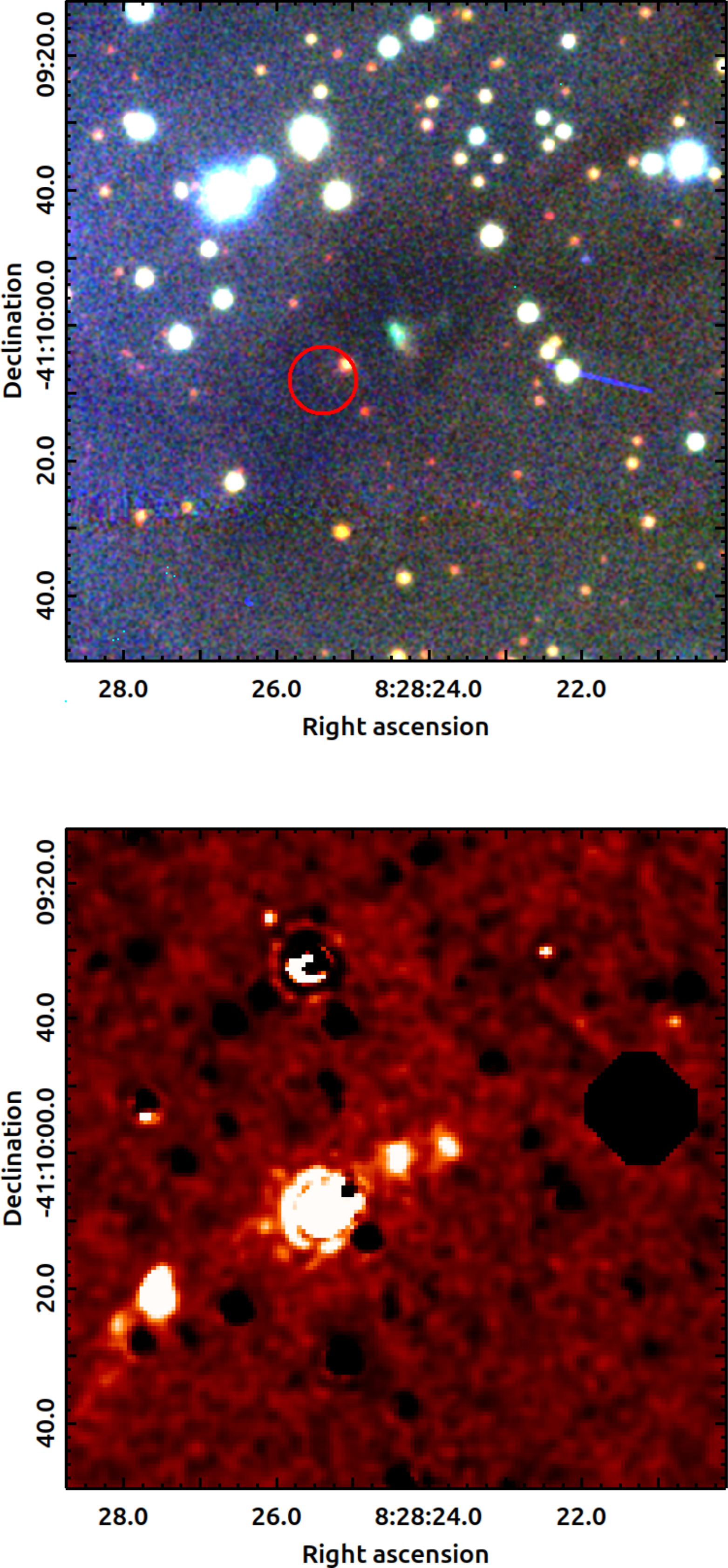}}
    \caption{{\bf Top:} Same as Fig.\,\ref{fig19} for Bres~20 (center). The red circle marks the position of the AllWISE source. {\bf Bottom:} The IRAC I2$-$I1 difference image reveals the full extent of the outflow traced by the line emission in the I2 channel. The black square is due to a region of missing values in the I1 image.
    }
    \label{fig20}
\end{figure}

This is another example of compact nebulosity centered on an anonymous dark cloud (Fig.\,\ref{fig20}). The presence of ALLWISE J082825.37-411008.1 aka IRAS~08266-4100 points to ongoing star formation.  The DECaPS2 $r$ image shows a clear flux excess at the position of Bres 20 (Fig.\,\ref{fig20} top), which indicates the presence of typical HHO emission lines. At this location, there is neither a 2MASS nor an AllWISE source, which underlines the HH nature.

The source J082825.37-411008.1 is by far the reddest and the brightest in the W4 band of all AllWISE sources covered by the dark cloud. It is driving a bipolar outflow in the south-east/north-west direction, as indicated by the IRAC I2$-$I1 difference image shown in the bottom panel of Fig.\,\ref{fig20}. The HHO is obviously a major shock of the blue-shifted flow, while the red-shifted one is hidden in the optical by the dark cloud.

\subsection{Bres 21}
\begin{figure}[!htbp]
\centerline{\includegraphics[width=.75\linewidth]{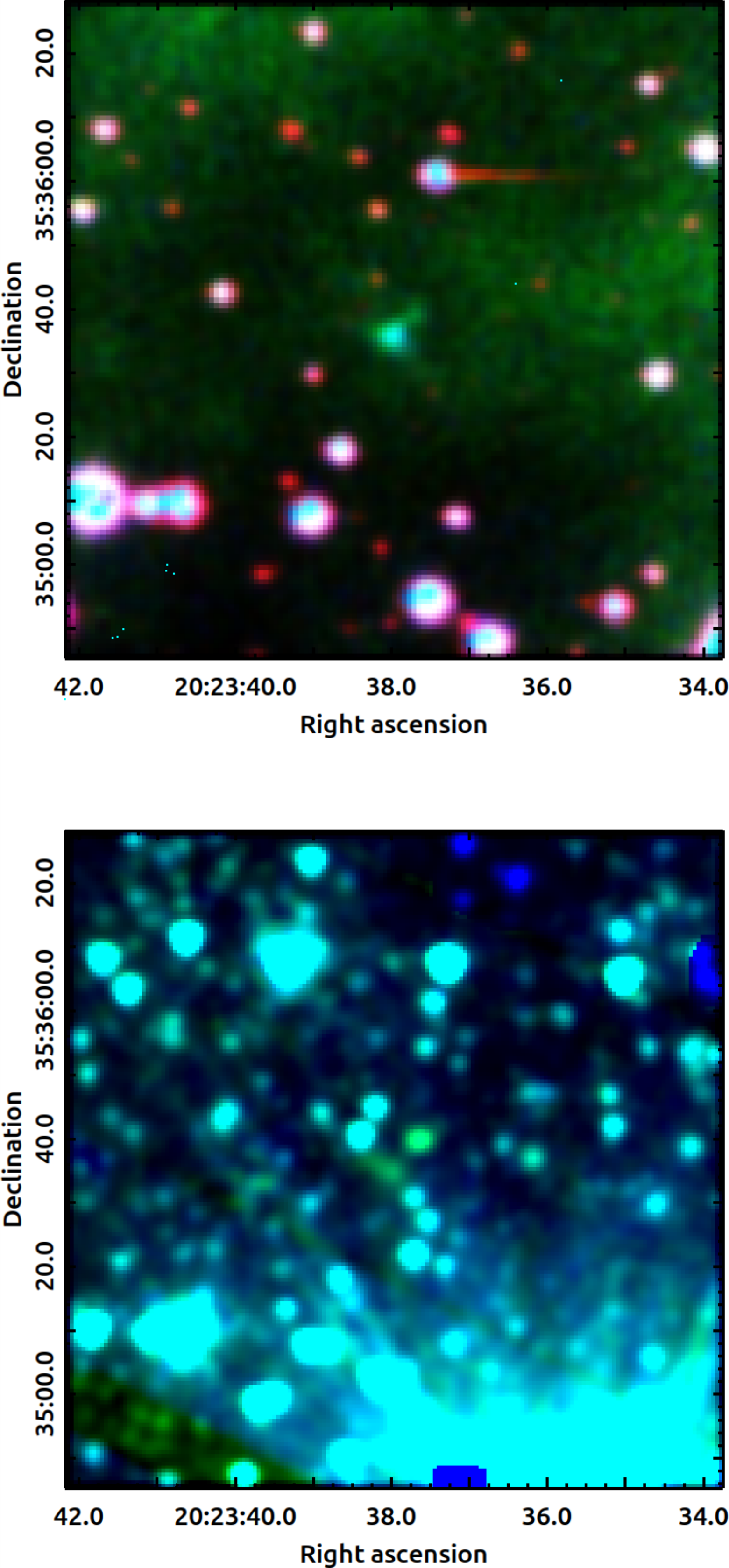}}
    \caption{{\bf Top:} Same as Fig.\,\ref{fig5} for Bres~21. The HHO becomes visible against the H$\alpha$ background due to strong [{S\sc{ii}}] emission. {\bf Bottom:} IRAC color image with I1 (blue) and I2 (green) shows compact excess emission in the I2 band coinciding with the HHO.
    }
    \label{fig21}
\end{figure}

This brightness patch (Fig.\,\ref{fig21}) is on top of widespread H$\alpha$ emission. Thus, its detection would not have been possible if another emission line, likely [{S\sc{ii}}], were not stronger than H$\alpha$. Bres~21 seems to consist of two components. This is confirmed by the IRAC I2 image, which shows the north-west component to be brighter. The exciting source for this HHO could not be identified. It may be possible that Bres~21 is a compact shock region in the ISM that belongs to a parsec-scale outflow.

\subsection{Bres 22}
\begin{figure}[!htbp]
\centerline{\includegraphics[width=.75\linewidth]{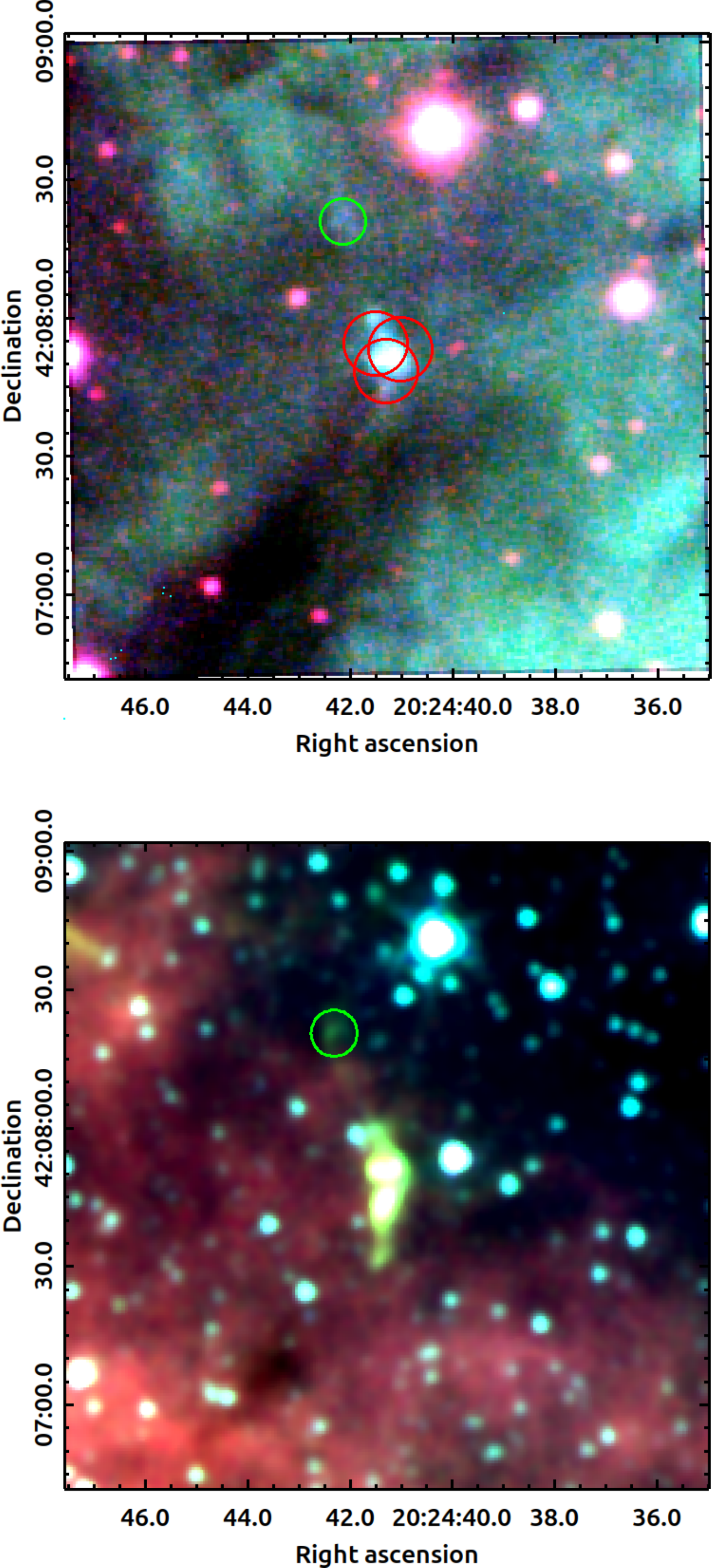}}
    \caption{{\bf Top:} Same as Fig.\,\ref{fig5} for Bres~22. The HHO becomes visible against the H$\alpha$ background due to strong [{S\sc{ii}}] emission. {\bf Bottom:} IRAC color image with I1 (blue) and I2 (green) shows compact excess emission close to the HHO.
    }
    \label{fig22}
\end{figure}

Similarly to Bres~21, Bres~22  (Fig.\,\ref{fig22}) is also seen against a variable background of an H{\sc ii} region, interspersed with dark clouds in the foreground. The ALLWISE catalog contains three close-by entries with similar magnitudes in each band, possibly due to dissection of extended emission. Simbad lists the object as a member of the ``2MASS-selected Flat Galaxy Catalog'', despite \cite{2019MNRAS.482.5167S} classifying it as ``obviously not a galaxy''. The latter is supported by our analysis, since it seems to drive an outflow in the north-south direction, which is bent. Some features can be identified with strong [{S\sc{ii}}] emission in the optical that coincides with patches in I2 frames.  Although the source is included by various photometric surveys, it lacks radio measurements of molecular lines, which, among others, would provide a kinematic distance estimate. Due to its extended shape, GAIA is not capable of performing astrometry for this object. 

\subsection{Bres 23}
\begin{figure}[!htbp]
\centerline{\includegraphics[width=.75\linewidth]{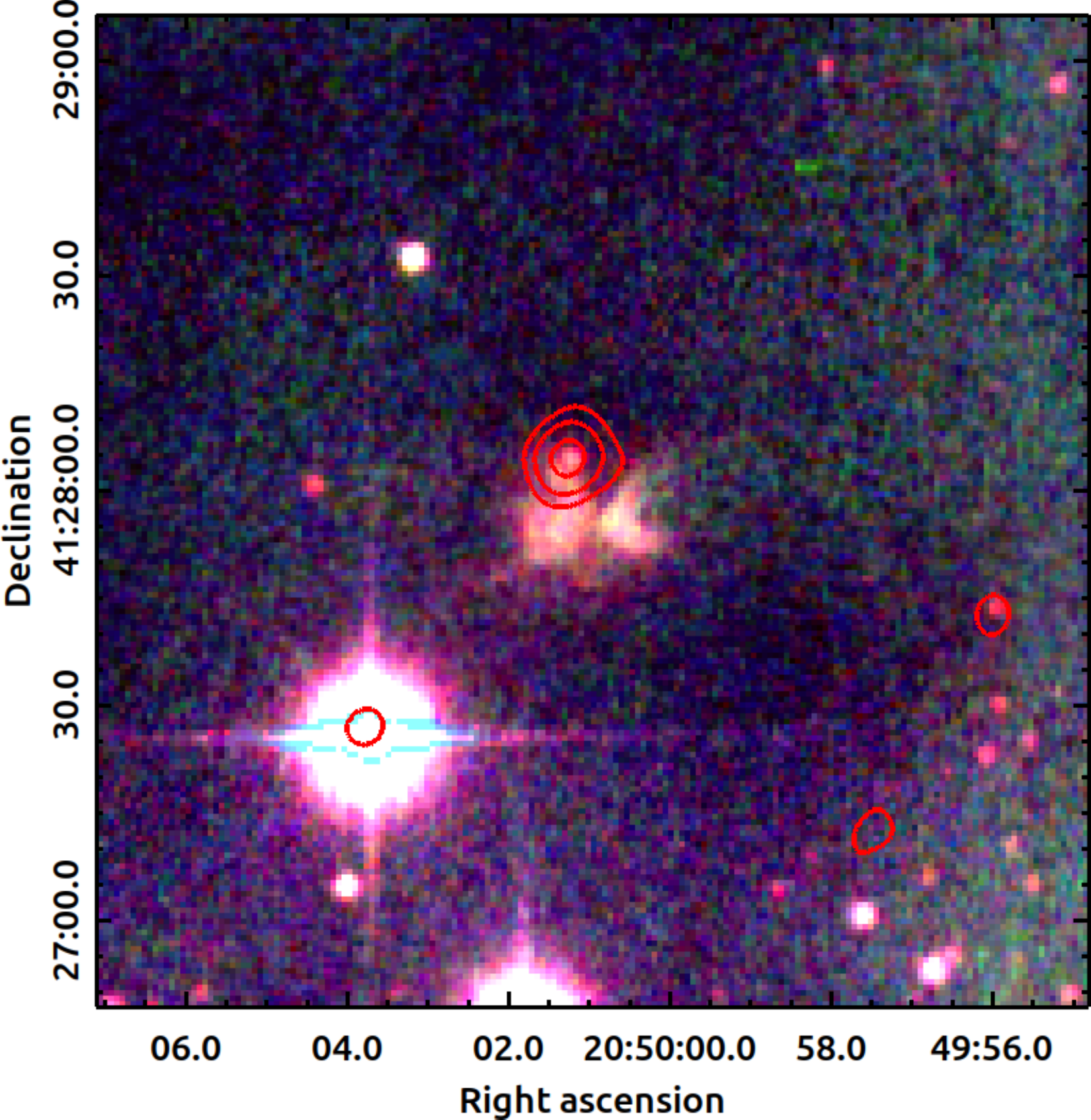}}
    \caption{ Same as Fig.\,\ref{fig5} for Bres~23. The red contours denote the W3 emission from the corresponding unWISE image.
    }
    \label{fig23}
\end{figure}

This object appeared to be a clear-cut case at first sight, since its biconical morphology  (Fig.\,\ref{fig23}) is typical for YSOs surrounded by an envelope seen edge-on. In such a case, the protostar is located at the tail behind its circumstellar disk. However, as Fig.\,\ref{fig23} indicates, the contours of the W3 emission are displaced with respect to this location. A deeply embedded very young protostar of Class 0 or I (\citealt{1993ApJ...406..122A}; \citealt{1984ApJ...287..610L}) might be undetected even at the W3 wavelength, but its bipolar cavities would hardly be visible in the optical. In addition, no signs of jet/outflow activity have been found. This might happen when a YSO is in quiescence in between accretion bursts. Although located in the dark cloud Dobashi~2635 \citep{2011PASJ...63S...1D}, the nebulosity Bres~23 does not qualify as HHO and its association with star formation is unclear. 

\subsection{Bres 24}
\begin{figure}[!htbp]
\centerline{\includegraphics[width=.75\linewidth]{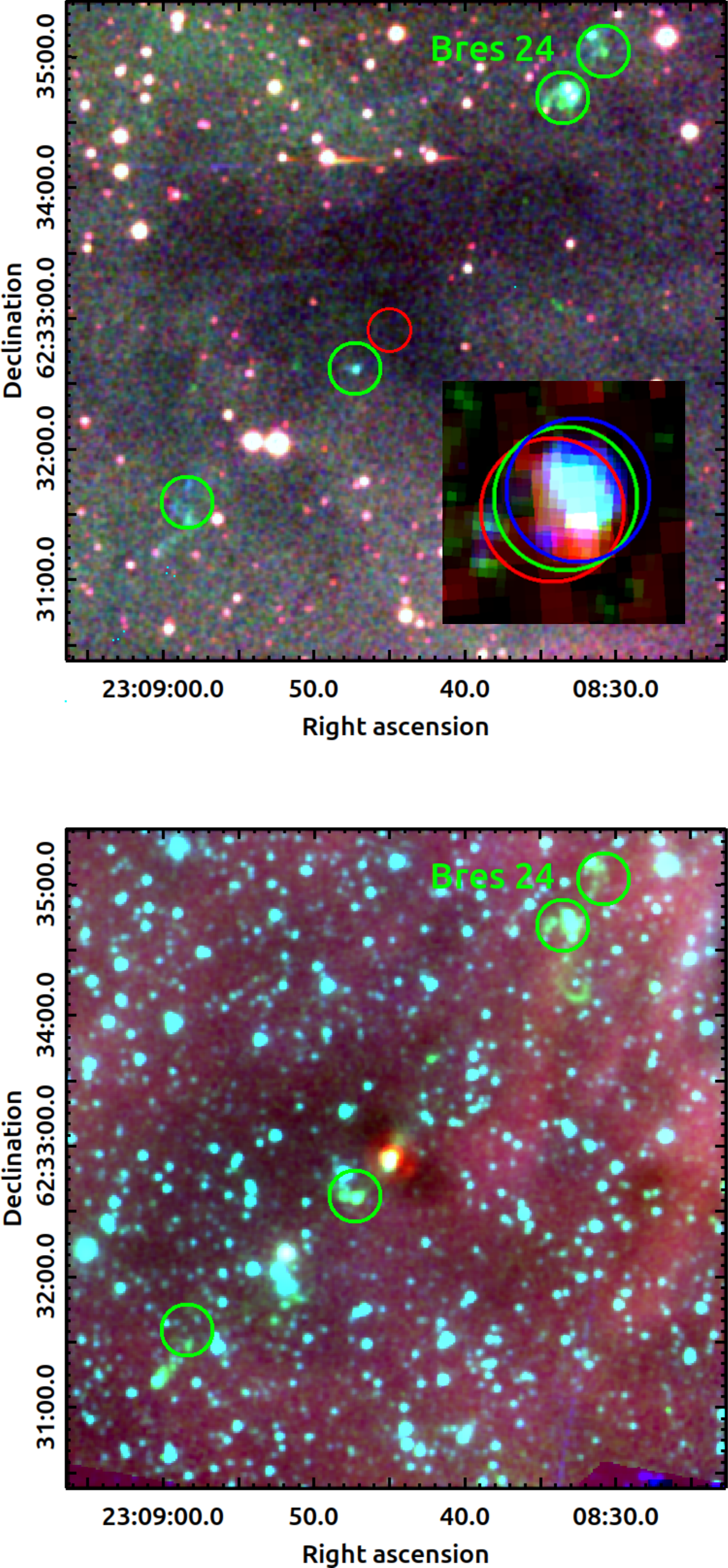}
}
    \caption{{\bf Top:} Same as Fig.\,\ref{fig5} for Bres~24 with marked HHOs. The zoom inset shows a color composite of  Bres~24 based on POSS-I (red), POSS-II (green), and TAUKAM H$\alpha$ (blue) which reveals its proper motion. The circles mark the centroid position of the enclosed emission. {\bf Bottom:} M1, I2, and I1 RGB composite.
    }
    \label{fig24}
\end{figure}

Bres~24 (Fig.\,\ref{fig24}, top) is visible in the R-band frames of POSS-I \citep{minkowski_abell_poss_i_1968} and POSS-II \citep{1991PASP..103..661R}. Even in 2MASS, there is a sign in the K band due to the emission of shocked H$_2$. It shows a binary morphology that, at first sight,  might be taken as evidence of a YSO seen close to edge-on. However, there is no infrared source at this position. 

Bres~24 is located north-west of an anonymous dark cloud. To the south-east of this cloud, knotty emission line features are visible in the TAUKAM narrowband images. This led to the suspicion that the driving source of a bipolar HH flow might be hidden in the dark cloud. The bottom panel of Fig.\,\ref{fig24} shows a color composite based on IRAC1 (blue), IRAC2 (green) and MIPS \citep{2004ApJS..154...25R} Band 1 (M1, 24\,\micron, red). It revealed that the YSO that gives rise to the flow is deeply embedded in the infrared dark cloud.  In fact, it harbors ALLWISE J230844.93+623254.3 which is the driving source. Most of the flow features seen in the 4.5\,\micron{} IRAC2 band have optical counterparts, which is rarely the case. The bowlike emission south of Bres~24, not aligned with the main flow, is likely due to another YSO that has not been identified yet.
 
The YSO, which has no Simbad entry so far, was observed by several mid- and far-IR as well as radio surveys. Analysis of 2MASS and Spitzer photometry pointed to a protostellar candidate that resides in the Cepheus OB3 association \citep{2012AJ....144...31K}. Using the radial velocity of the CO(1$-$0) emission line, a kinematic distance of $\sim$1\,kpc was derived \citep{2021A&A...646A..74M}. At this distance, the project length of the outflow amounts to $\sim$1.5\,pc which promotes it to be a parsec-scale flow. 
The inset of Fig.\,\ref{fig24} shows a zoom on Bres~24 in RGB representation composed of the POSS-I, POSS-II and TAUKAM H$\alpha$ frames, which have the epochs of 1953, 1991, and 2024. It clearly shows the proper motion of the bow shock away from the YSO. From the relative angular displacement of the centroids of the enclosed emission, a tangential velocity of 177.4$\pm$3.0\,km\,s$^{-1}$ was derived. This corresponds to a dynamic age of $\sim$66\,kyr. 

\subsection{Bres 25}
\begin{figure}[!htbp]
\centerline{\includegraphics[width=.75\linewidth]{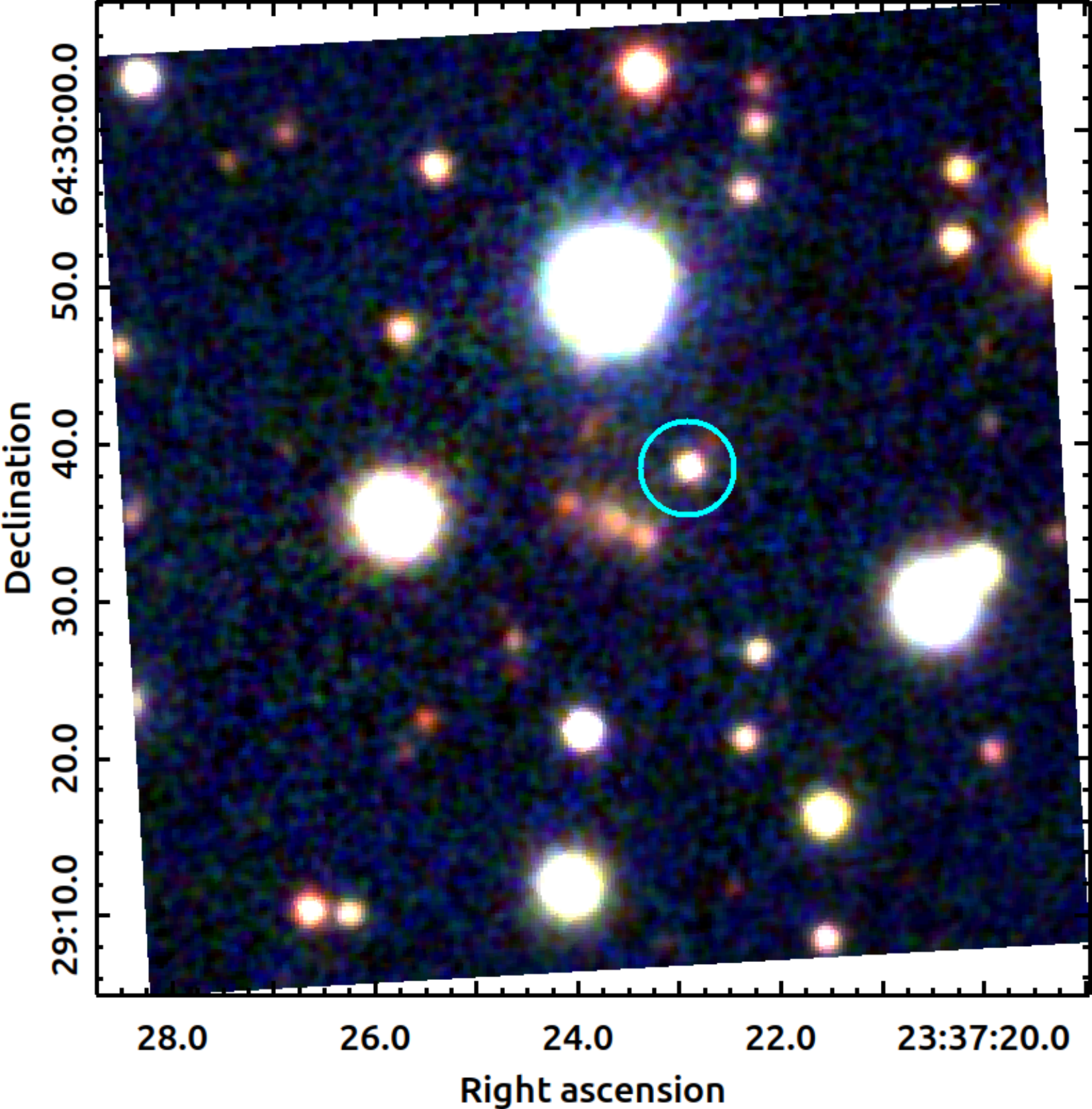}}
    \caption{{\bf Top:} Same as Fig.\,\ref{fig1} for Bres~25. The cyan circle marks the star for which its distance has been quoted.
    }
    \label{fig25}
\end{figure}

This object consists of three nebulous knots (Fig.\,\ref{fig25}). One of those has an H$\alpha$ excess of $\sim$0.5\,mag according to IPHAS \citep{2005MNRAS.362..753D}. \cite{2016MNRAS.455.3126C} argue that the object is an embedded cluster. This is supported by extended emission in the WISE W3 and W4 bands, which arises from polycyclic aromatic hydrocarbons due to UV excitation by OB stars \citep{1989ApJS...71..733A}. Far-infrared emission from warm dust is also present. The lack of an associated dark cloud in the optical suggests that the cluster is fairly distant. The star closest to the object observed by GAIA has a photogeometric distance of 4.5\,kpc, which seems to support this assumption. Rather than HHOs, the knots may represent scattered light from OB stars, with the H$\alpha$ excess arising from an associated H{\sc ii} region.


\section{Conclusions}\label{sec5}
The present work represents the continuation of a citizen science search for compact optical emission nebulae that indicate both young and mature stars. It started with the discovery of Bres 1 \citep{2023RNAAS...7..254B}, a YSO with Herbig-Haro features. Since then, 24 new candidates have been identified with the help of the Aladin interactive sky atlas \citep{2000A&AS..143...33B} in publicly available optical sky surveys. To pin down their characteristics, narrowband imaging was performed for a subsample of 14 objects using TAUKAM at the AJT.

The vast majority of the compact nebulosities found are associated with YSOs; many of them have not been studied yet. This underlines the diagnostic power of our method in identifying this class of astrophysical objects. With the help of the deep narrowband imaging and the use of complementary IR imagery, fainter HH candidates have been revealed. The majority belong to jet-driven bipolar HH flows. 

In view of the accretion–ejection connection observed in YSOs (e.g., \citealt{2018A&A...612A.103C}), the fact that some sources (Bres 13, 14, 19, 23) do not exhibit HHOs suggests that they may currently be in a quiescent phase with low accretion rates. This underlines that disk-mediated accretion of YSOs is a non-steady, episodic process (e.g., \citealt{2024AJ....167...72D}).

Moreover, seemingly isolated HH candidates have been found (Bres 12, 21) for which exciting sources could not be identified. These may be part of parsec-scale flows with driving sources located at considerable distance. Proper motion studies are required to get a better idea of where those objects came from.



Our citizen science project shows that this collaboration between experts and amateurs is extremely valuable. In order to optimize and automate the time-consuming manual search in the Aladin interactive sky atlas \citep{2000A&AS..143...33B}, the coauthor is developing AI-based search methods, which were not yet fully completed at the time of publication. A corresponding publication will follow.

\section*{Acknowledgments}
This research has made use of the SIMBAD database, operated at CDS, Strasbourg, France, and the NASA/IPAC Infrared Science Archive, which is funded by the National Aeronautics and Space Administration and operated by the California Institute of Technology.
This publication makes use of data products from the Wide-field Infrared Survey Explorer, which is a joint project of the University of California, Los Angeles, and the Jet Propulsion Laboratory/California Institute of Technology, funded by the National Aeronautics and Space Administration.
This research has made use of the VizieR catalogue access tool, CDS,
 Strasbourg, France (DOI : 10.26093/cds/vizier). The original description 
 of the VizieR service was published in 2000, A\&AS 143, 23.






\bibliography{Wiley-ASNA}%

\newpage
\appendix
\section{Full Table of objects}

\renewcommand{\tabcolsep}{3pt}
\begin{table}[h]
\centering
\caption{}
\label{tab:coords}
\begin{tabular}{|c|m{1.5cm}|m{1.45cm}|m{.2cm}|c|l|}
\hline
{Bres ID} & RA & Dec & T &HH&Simbad Identifier\\
\hline
 2 & 00 09 46.0 & +65 33 36 & &&\\              
 3 & 00 12 51.1 & +60 29 29 & &&\\             
 4 & 00 57 20.0 & +62 39 55 & &&ZOAG123.61-0.20 \\               
 5 & 01 20 03.0 & +66 51 14 & + & + &HH1227\\       
 6 & 03 22 36.0 & +63 32 07 & + &   &2MASX 03223621+633205\\
 \hphantom{0}6A& 03 22 23.8 & +63 30 53 & + &+& \\
 \hphantom{0}6B  & 03 22 29.3 & +63 30 31 & + &+& \\
 7 & 03 54 37.2 & +53 12 36 & + & +&IRAS 03510+5301\\             
 \hphantom{0}7A  & 03 54 47.7 & +53 10 36 & + &+&\\
 \hphantom{0}7B  & 03 55 01.9 & +53 08 29 & + &+&\\
 \hphantom{0}7C  & 03 55 07.2 & +53 07 26 & + &+&\\
\hphantom{0} 7D  & 03 55 12.0 & +53 09 25 & + &+&\\ 
 8 & 04 30 49.0 & +63 46 13 & + & + &IRAS 04261+6339\\             
 9 & 04 36 39.0 & +54 36 55 &   & + &HH378A \\       
10\hphantom{0} & 04 43 38.6 & +41 09 41 & + &   &IRAS 04401+4103\\ 
10A   & 04 43 38.3 & +41 09 35 & + &+&\\
11\hphantom{0} & 05 36 30.3 & +31 49 14 &+& &\\           
12\hphantom{0} & 05 36 23.0 & +31 45 51 &+& &2MFGC 4557\\               
12A  & 05 36 43.4 & +31 44 23 &+&+& \\
13\hphantom{0} & 05 44 26.1 & +21 29 14 & +  && \\             
14\hphantom{0} & 05 44 28.0 & +21 30 09 & + & & \\             
15\hphantom{0} & 05 46 21.0 & +26 28 49 & + & &\\             
15A   & 05 46 17.3 & +26 29 41 & + &+&\\
16\hphantom{0} & 06 08 42.0 & +29 36 43 &   &+&\\               
16A   & 06 08 41.7 & +29 37 03 &   &+&\\
17\hphantom{0} & 06 11 57.0 & +20 11 36 &   & + &\\               
18\hphantom{0} & 06 38 29.0 & $-$00 57 42 & && IRAS 06359-0055\\ 
18A   & 06 38 32.0 & $-$00 57 45 & &+ &\\
18B   & 06 38 22.9 & $-$00 58 07 & &+ &\\
19\hphantom{0} & 08 26 12.2 & $-$41 10 30 & & & \\
20\hphantom{0} & 08 28 24.4 & $-$41 10 01 & &+& \\            
21\hphantom{0} & 20 23 38.0 & +35 35 36 &+& + &\\             
22\hphantom{0} & 20 24 41.0 & +42 07 51 &+& + &2MFGC15507 \\    
22A   & 20 24 42.4 & +42 08 21 & &+&\\
23\hphantom{0} & 20 50 00.0 & +41 27 56 &+& &\\               
 1 & 20 53 25.0 & +51 08 17 & & + &2MASS J20532531+5108186\\             
24\hphantom{0} & 23 08 33.1 & +62 34 43 &+& + &\\               
24A   & 23 08 30.7 & +62 35 02 &+&+&\\
24B   & 23 08 47.2 & +62 32 36 &+&+&\\
24C   & 23 08 58.4 & +62 31 35 &+&+&\\
25\hphantom{0} & 23 37 23.0 & +64 29 34 & & &\\               
\hline
\end{tabular}
\end{table}

\end{document}